\pdfoutput=1
\documentclass[sigconf, screen, nonacm]{acmart}

\usepackage{textcomp}
\usepackage{booktabs}
\usepackage{array}
\usepackage{enumitem}
\usepackage{xspace}
\usepackage{tikz}
\usetikzlibrary{arrows.meta,calc,decorations.pathreplacing}

\newcommand{\sys}{Contexture\xspace}

\begin{document}
\begin{sloppy}

\title{\sys: Runtime Context Eviction for High-Throughput Agent Serving}

\author{Colin Hong}
\affiliation{%
  \institution{Nanyang Technological University}
  \country{Singapore}}
\author{Bryan Shan}
\affiliation{%
  \institution{Nanyang Technological University}
  \country{Singapore}}
\author{Wei Siong Pang}
\affiliation{%
  \institution{Nanyang Technological University}
  \country{Singapore}}
\author{Chengzhi Lu}
\affiliation{%
  \institution{Nanyang Technological University}
  \country{Singapore}}
\author{Xu Guo}
\affiliation{%
  \institution{Nanyang Technological University}
  \country{Singapore}}
\author{Nitin Shivaraman}
\affiliation{%
  \institution{IAIC, A*STAR}
  \country{Singapore}}
\author{Tao Luo}
\affiliation{%
  \institution{IAIC, A*STAR}
  \country{Singapore}}
\author{Esha Choukse}
\affiliation{%
  \institution{Microsoft}
  \country{USA}}
\author{Kisson Lin}
\affiliation{%
  \institution{SuperMem Pte. Ltd.}
  \country{Singapore}}
\author{Dmitrii Ustiugov}
\affiliation{%
  \institution{Nanyang Technological University}
  \country{Singapore}}
\renewcommand{\shortauthors}{Hong et al.}

\begin{abstract}
LLM agents such as Hermes, OpenClaw, and Codex repeatedly resend growing execution histories, increasing shared GPU key-value (KV) cache demand in high-bandwidth memory (HBM). We show that shared KV capacity limits agent-serving throughput. Existing context-reduction methods lack global cache-state visibility, while runtime KV managers lack tool-execution information. Tracing tokens to their originating tool operations reveals that later operations supersede earlier outputs and associated turns, identifying 64\% of recorded final-history tokens as removal candidates.

We present \sys, a harness- and model-agnostic architecture for runtime context eviction. \sys separates the harness's full conversation history from the model's execution context, reducing subsequent requests' KV demand without modifying harness logic or model execution. It combines tool-execution information with global visibility into shared KV pressure and prefix residency: supersession identifies removal candidates, while runtime state determines when removal justifies sacrificing cached computation. This coordination reduces memory demand and decode KV traffic while limiting re-prefill interference with concurrent decoding, without additional model calls or training.
Across two models, two harnesses, and two benchmarks, \sys improves throughput under SLO by 20--73\% over the best of no eviction, harness compaction, AgentDiet, and Continuum, and by 7--63\% over observation masking in five of eight configurations, and reduces GPU energy per completed session by up to 11\%. It also raises throughput under SLO by 23\% over LMCache alone on H100. Live-agent experiments show a mean accuracy improvement of 1.4 percentage points over unloaded no eviction, with a worst-case loss of 3.7 points, while observation masking loses 3--28 points.
\end{abstract}

\maketitle

\section{Introduction}
\label{sec:intro}

LLM agents such as Hermes~\cite{nous2026hermes}, OpenClaw~\cite{openclaw2026compaction}, and Codex~\cite{openai2026codexautocompact} perform multi-step tasks by repeatedly inspecting files, executing commands, and processing the resulting observations with LLMs. These interactions accumulate into a session history that is carried into subsequent model requests and can amount to hundreds of thousands of tokens~\cite{park2026rethinking,anthropic2026claudecodemodelconfig}.
To avoid recomputing growing session histories, the serving runtime retains their reusable KV prefix caches in shared GPU high-bandwidth memory (HBM), where concurrent sessions compete for finite cache capacity.

Our key observation is that shared KV capacity is the critical bottleneck that limits agent-serving throughput. As more sessions arrive, the session completion rate stops increasing once the available KV cache capacity is exhausted. Beyond this point, the runtime preempts requests and recomputes their KV caches, consuming additional compute without increasing completed-session throughput. Halving the KV pool of an H200 GPU reduces peak throughput by about 40\%. These observations motivate reducing the KV memory occupied by each session to increase serving capacity.

Existing approaches address growing KV memory demand at different levels of the agent-serving stack. At the \emph{harness level}, compaction~\cite{nous2026hermescompression,openclaw2026compaction} and LLM-driven history editing~\cite{xiao2026agentdiet} shorten session histories, but operate with a \emph{local view} that lacks information about shared memory pressure and cached KV residency. 
Their model judgments consume GPU time and energy, delaying session completion. They may also remove information the agent still needs.
 At the \emph{model level}, KV pruning selects tokens based on {attention scores within a request}~\cite{zhang2023h2o,li2024snapkv,cai2025rkv} {or across a session's turns~\cite{li2026intentkv}}, while other approaches require {training} to identify obsolete history~\cite{kariyappa2026sidequest,zhang2026memory,wang2026swe}. These methods do not account for shared KV memory pressure when deciding what to remove. At the \emph{runtime level}, KV cache management has a \emph{global view} of shared memory~\cite{li2025continuum,fang2026saecache}, but cannot shorten the session history, so evicted KV may be regenerated on subsequent requests. These approaches leave a coordination gap between identifying removable history and deciding when its removal is worthwhile given shared KV pressure and cached-prefix residency.

Closing this gap requires identifying historical context that later agent operations have superseded. We characterize redundancy in the KV cache by tracing historical tokens to their origins in agent tool executions, such as file reads and command executions. We find that later tool operations frequently supersede earlier observations of the same state, making the associated history redundant. Such supersession identifies 64\% of the tokens in the final histories of our recorded sessions as removal candidates, without additional LLM calls.
Each request exposes these candidates to the runtime, but removing them raises two challenges. A superseded observation may still support ongoing agent actions. Removing an earlier span also invalidates subsequent cached KV, introducing re-prefill that interferes with concurrent decoding. Eviction must therefore balance reduced memory demand against lost cache reuse.

\noindent\textbf{Our work.} We present \sys, a harness- and model-agnostic architecture for runtime context eviction that combines \emph{origin-based context eviction} with \emph{global cache-aware eviction}. \sys separates identifying removal candidates from deciding when to commit them. It uses tool operations and their targets to identify superseded outputs and associated turns, without additional model inference. Candidates remain available for a configurable grace window, giving the agent additional rounds to complete related actions before removal.

For each request, \sys combines eligible candidates with the runtime's global view of shared KV pressure and cached-prefix residency across sessions. If no cached KV follows the earliest removal point, it commits immediately: the next request would need to recompute that suffix anyway, so removal reduces work without sacrificing cache reuse. Otherwise, \sys defers removal until memory pressure warrants freeing capacity and limits the cached computation invalidated relative to the context removed. This coordination makes eviction responsive to both the demand from concurrent sessions and the cost of losing reusable computation.

The harness retains its full conversation history, while the runtime filters the execution context processed by the model. The shortened prompt uses the engine's ordinary prefix reuse and re-prefill, preserving exact execution over the retained context. This separation reduces subsequent KV demand and decoding work without modifying harness logic or model execution, or requiring additional model calls or training.

We implement \sys in vLLM~\cite{kwon2023efficient} and evaluate eight configurations spanning two models, two harnesses, and two benchmarks on H100 and H200 GPUs. \sys improves throughput under SLO by 20--73\% over the best-performing baseline among no eviction, default and tuned compaction, AgentDiet, and Continuum. At the corresponding baseline SLO throughput, \sys reduces p90 session slowdown by 30--78\%. Compared with observation masking, \sys achieves 7--63\% higher throughput in five configurations; masking performs better in the other three, all with Qwen MoE, but incurs 3--28 percentage points of accuracy loss across all configurations. With Qwen MoE and Hermes on Terminal-Bench, \sys retains its gains with larger HBM capacity and KV offloading, improving throughput by 14\% over no eviction on H200 and 23\% over LMCache alone on H100. It also reduces GPU energy per completed session by 8--11\% at high load. Live-agent experiments across eight configurations and three loads show a mean accuracy gain of 1.4 percentage points over unloaded no eviction, with a worst-case loss of 3.7 points.
In summary, we make the following contributions:
\begin{itemize}[leftmargin=0cm,itemindent=.2cm,labelwidth=\itemindent,labelsep=0cm,align=left]
\setlength\itemsep{0pt}
  \item We identify a fundamental trade-off in runtime context eviction: reducing memory demand can destroy reusable KV computation and stall concurrent decoding, making shared memory pressure and cache residency essential to deciding when to evict.
  \item We reveal that tool-operation supersession identifies 64\% of tokens in recorded final histories as removal candidates, providing a selection signal without additional \mbox{model inference}.
  \item We design \sys, a harness- and model-agnostic architecture for runtime context eviction that combines execution-based selection with global visibility into shared KV pressure and cache residency, balancing memory savings against recomputation and interference with concurrent decoding.
  \item We show throughput and energy benefits from \sys while improving task accuracy on average, with throughput gains persisting alongside larger HBM capacity and KV offloading.
\end{itemize}

\section{Background: Agentic Stack and Workload}
\label{sec:background}
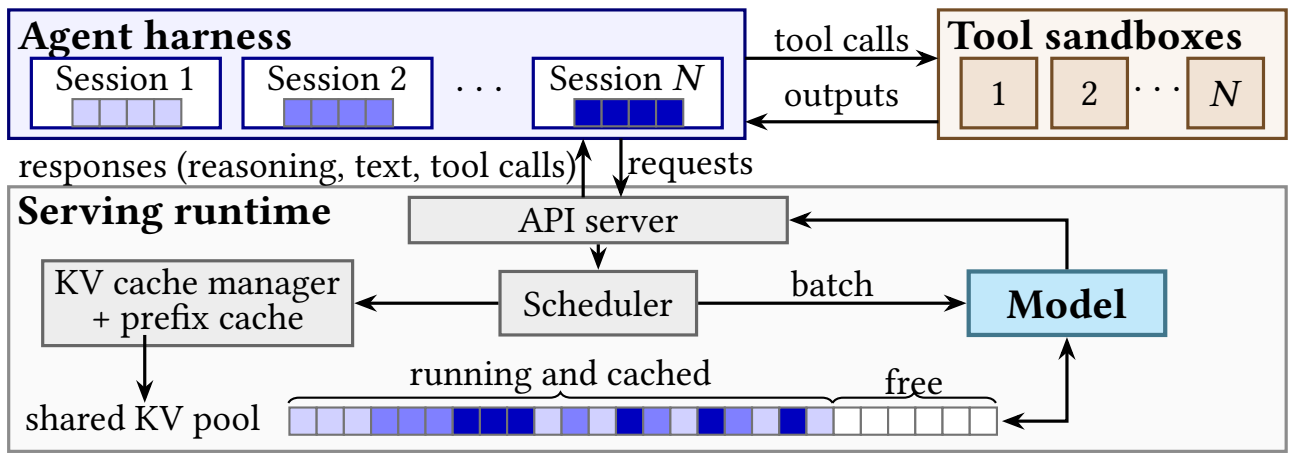
\begin{figure}[t]
  \centering
  \usetikzlibrary{arrows.meta,calc,decorations.pathreplacing}%
\begin{tikzpicture}[
  x=1cm, y=1cm,
  font=\fontsize{7}{8}\selectfont,
  hlayer/.style={draw=blue!45!black, semithick, fill=blue!5},
  tlayer/.style={draw=brown!60!black, semithick, fill=brown!8},
  rlayer/.style={draw=black!45, semithick, fill=gray!4},
  sess/.style={draw=blue!55!black, semithick, fill=white},
  sbox/.style={draw=brown!60!black, semithick, fill=brown!22, align=center, inner sep=1.5pt},
  box/.style={draw=black!60, semithick, fill=gray!14, align=center, inner sep=1.5pt},
  mbox/.style={draw=cyan!45!black, thick, fill=cyan!22, align=center, inner sep=1.5pt},
  lyr/.style={font=\fontsize{8}{9.6}\selectfont\bfseries, anchor=north west, inner sep=1.5pt},
  small/.style={font=\fontsize{7}{7.8}\selectfont},
  flow/.style={-{Stealth[length=1.5mm]}, semithick},
  cell/.style={draw=black!55, thin, minimum size=1.8mm, inner sep=0pt, anchor=west, fill=white},
  sA/.style={cell, fill=blue!18},
  sB/.style={cell, fill=blue!50},
  sN/.style={cell, fill=blue!75!black},
  brc/.style={decorate, decoration={brace, amplitude=2pt}, thin},
]
\def\cw{0.18}%

\draw[hlayer] (0,2.07) rectangle (4.87,2.92);
\draw[tlayer] (6.15,2.07) rectangle (8.45,2.92);
\draw[rlayer] (0,0) rectangle (8.45,1.75);
\node[lyr] at (0,2.92) {Agent harness};
\node[lyr] at (6.15,2.92) {Tool sandboxes};
\node[lyr] at (0,1.75) {Serving runtime};

\foreach \x/\n/\s in {0.15/1/sA, 1.55/2/sB, 3.47/N/sN} {
  \draw[sess] (\x,2.14) rectangle ++(1.25,0.46);
  \node[inner sep=0pt] at (\x+0.625,2.47) {Session $\n$};
  \foreach \i in {0,...,3} \node[\s] at ({\x+0.265+\i*\cw},2.24) {};
}
\node at (3.135,2.37) {$\cdots$};

\foreach \x/\n in {6.3/1, 6.9/2, 7.8/N} \node[sbox, minimum width=0.5cm, minimum height=0.46cm] at (\x+0.25,2.37) {$\n$};
\node at (7.62,2.37) {$\cdots$};
\draw[flow] (4.87,2.6) -- node[small, above=0.5pt, inner sep=0.5pt] {tool calls} (6.15,2.6);
\draw[flow] (6.15,2.18) -- node[small, above=0.5pt, inner sep=0.5pt] {outputs} (4.87,2.18);

\node[box, minimum width=2.5cm, minimum height=0.3cm] (api) at (3.9,1.53) {API server};
\draw[flow] (3.8,1.68) -- (3.8,2.07);
\draw[flow] (4.05,2.07) -- (4.05,1.68);
\node[small, anchor=east, inner sep=1pt] at (3.8,1.875) {responses (reasoning, text, tool calls)};
\node[small, anchor=west, inner sep=1pt] at (4.05,1.875) {requests};
\node[box, minimum width=2.05cm, minimum height=0.42cm] (kvm) at (1.25,0.98) {KV cache manager\\[-1.5pt]+ prefix cache};
\node[box, minimum width=1.3cm, minimum height=0.42cm] (sched) at (3.9,0.98) {Scheduler};
\node[mbox, minimum width=1.3cm, minimum height=0.42cm] (llm) at (7.0,0.98) {\fontsize{8}{9.6}\selectfont\bfseries Model};
\draw[flow] (api) -- (sched);
\draw[flow] (sched) -- (kvm);
\draw[flow] (sched) -- node[small, above=0.5pt, inner sep=0.5pt] {batch} (llm);
\draw[flow] (llm.north) |- (api.east);

\def\px{1.85}\def\py{0.2}
\node[anchor=west, inner sep=0pt] (pool) at (0.1,\py) {shared KV pool};
\foreach \i/\s in {0/sA,1/sA,2/sA,3/sB,4/sB,5/sB,6/sN,7/sN,8/sN,9/sA,10/sB,11/sA,12/sN,13/sB,14/sA,15/sN,16/sB,17/sA,18/sN,19/sA} \node[\s] at ({\px+\i*\cw},\py) {};
\foreach \i in {20,...,25} \node[cell] at ({\px+\i*\cw},\py) {};
\draw[flow] (0.9,0.77) -- (pool.north -| 0.9,0);
\draw[flow, {Stealth[length=1.5mm]}-{Stealth[length=1.5mm]}] ({\px+26*\cw+0.04},\py) -| (llm.south);
\draw[brc] (\px,{\py+0.12}) -- node[above=1.5pt, inner sep=0pt] {running and cached} ({\px+20*\cw},{\py+0.12});
\draw[brc] ({\px+20*\cw},{\py+0.12}) -- node[above=1.5pt, inner sep=0pt] {free} ({\px+26*\cw},{\py+0.12});
\end{tikzpicture}%
  \vspace{-2em}
  \caption{The agentic serving stack.}
  \vspace{-2em}
  \Description{Two stacked boxes. Top: the agent harness, one instance per session, each holding the session's history. Requests from all sessions go down to the serving runtime, and responses (reasoning, text, tool calls) come back up. Bottom: the serving runtime. Requests enter at its API server and go to the scheduler, which allocates KV blocks through the KV cache manager with its prefix cache and runs the model; the model reads and writes a shared GPU KV pool whose blocks belong to several sessions or are free, and its output returns through the API server. Annotations: the harness has a local view (one session); the runtime has a global view (all sessions, the pool, cached prefixes); the model, inside the runtime, has a local view (one request).}
  \label{fig:agent-stack}
\end{figure}

\label{sec:bg-stack}
\label{sec:agent-context}

Figure~\ref{fig:agent-stack} illustrates the agent-serving stack, where an agent harness interacts with an LLM through a serving runtime. The harness, such as Hermes~\cite{nous2026hermes}, OpenClaw~\cite{openclaw2026compaction}, or Codex~\cite{openai2026codexautocompact}, sends model requests to a serving runtime such as vLLM~\cite{kwon2023efficient}. The runtime serves requests from multiple agent sessions on shared GPUs. Its API server receives incoming requests, the scheduler batches them for model execution, and the KV cache manager allocates KV blocks from a shared GPU memory pool. The model processes each request and produces reasoning, text, and tool calls conditioned on its history. The harness executes returned tool calls, such as shell commands, file reads, and file writes, in the session's tool sandbox and incorporates their outputs into the conversation history. We define a model request together with its ensuing tool execution as a \emph{round}, and the sequence of rounds belonging to one task as a \emph{session}. Each session runs its own harness instance.

Across successive rounds, each request carries the session's accumulated conversation history, which we call its \emph{logical context}. Consecutive requests share an increasingly long prefix. With prefix caching, the KV cache manager retains the KV of processed tokens in GPU memory after a request finishes, allowing the next round to prefill only its new suffix. This reuse depends on the preceding context remaining unchanged and its KV states remaining cached. Removing an earlier context span changes the prefix of subsequent tokens, invalidating their cached KV states and requiring the surviving suffix to be prefilled again. These KV blocks occupy a finite pool shared by concurrent sessions and remain idle while the harness executes tools. As sessions progress, their growing histories increase demand for this shared pool. When the pool fills, the runtime evicts cached blocks or preempts running requests, causing later requests to recompute KV states that are no longer resident.

\section{Agent Context and KV Management \& Limitations of Existing Approaches}
\label{sec:limitations}
\label{sec:motivation}

We review context and KV management (\S\ref{sec:bg-approaches}), establish the shared-KV capacity bottleneck (\S\ref{sec:obs-wall}) and examine limitations of three existing methods (\S\ref{sec:design-space}). We replay recorded sessions of Hermes~\cite{nous2026hermes} on Terminal-Bench~2.1~(TB2.1)~\cite{merrill2026terminal} at increasing load, served by vLLM~\cite{kwon2023efficient} with Qwen3.8-27B~\cite{qwen2026qwen38} (\S\ref{sec:eval-setup})%
. \emph{Throughput} is the number of sessions completed within their task budgets per minute, and \emph{throughput under SLO} is the throughput where their p90 slowdown relative to each task's unloaded no-eviction run reaches 10$\times$.

\subsection{Managing Agent Context and KV}
\label{sec:bg-approaches}

Management of agent context and KV states spans three levels of the serving stack: the harness, model, and serving runtime. Each level operates on different states and has a distinct scope of control.

\noindent\textbf{Harness.} The harness constructs the logical context for each round from the session's conversation history, deciding which earlier messages and tool outputs the model sees. It can shorten this history in several ways. Compaction replaces older history with an LLM-generated summary~\cite{nous2026hermescompression,openclaw2026compaction,openai2026codexautocompact,anthropic2026claudecodehow}; observation masking~\cite{lindenbauer2025complexity} replaces older tool outputs with placeholders while retaining the agent's reasoning and actions; and LLM-driven history editing asks a model which parts to remove~\cite{xiao2026agentdiet}. These methods reduce subsequent requests' context length and KV demand.

\noindent\textbf{Model.} Given the context assembled by the harness, model-level management uses information derived from model computation or training to identify tokens and their corresponding KV states for retention or removal. Representative examples include attention-based KV pruning and learned context selection~\cite{zhang2023h2o,li2024snapkv,cai2025rkv,kariyappa2026sidequest}. These methods make selection decisions at the token level within individual requests.

\noindent\textbf{Runtime.} Once the model has computed the KV states, the serving runtime manages their residency across concurrent sessions. It determines which cached states remain in GPU memory, are evicted, or are moved to other memory tiers. Representative mechanisms include KV eviction, offloading, and retention~\cite{fang2026saecache,feng2025evicpress,li2025continuum}. Offloading preserves cached computation outside GPU memory but leaves future context demand unchanged. We evaluate its combination with context eviction in \S\ref{sec:eval-offload}.

\subsection{KV Capacity Bounds Agent Serving}
\label{sec:obs-wall}

\begin{figure}[t]
  \centering
  \includegraphics[width=\columnwidth]{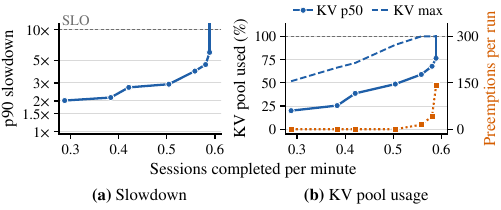}
\vspace{-2em}
\caption{P90 session slowdown, KV pool usage, and request preemptions versus throughput under increasing load without context eviction. (a) P90 session slowdown. (b) KV pool occupancy and preemptions per run.}
\vspace{-1em}
  \Description{Panel a shows p90 slowdown rising to 6.0 times at the peak throughput of 0.59 sessions completed per minute, under the 10 times SLO. Panel b shows peak KV occupancy reaching 100 percent at 0.56 sessions per minute, followed by increasing preemptions, reaching 140 per run at the peak throughput.}
  \label{fig:wall}
\end{figure}

We examine how shared KV capacity affects agent-serving throughput as the load increases.
Fig.~\ref{fig:wall} shows that without context eviction, throughput rises to 0.59 sessions/min before saturating, while p90 slowdown reaches 6.0$\times$. The throughput peak coincides with KV pool saturation, which first occurs at 0.56 sessions/min. As the load increases, preemptions rise to 140 per run, causing repeated KV recomputation without increasing throughput. To further examine the impact of KV capacity, we reduce the KV pool from 1.5M to 1.0M and 0.76M tokens on the same H200. Peak throughput drops to 0.45 and 0.35 sessions/min, respectively. These results establish shared KV capacity as a limiting factor for agent-serving throughput.

\subsection{Existing Methods Fall Short}
\label{sec:design-space}

\begin{figure}[t]
  \centering
  \includegraphics[width=\columnwidth]{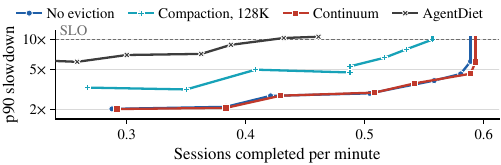}
\vspace{-2em}
  \caption{P90 session slowdown versus throughput under increasing load for no eviction, compaction, AgentDiet, and Continuum. The dashed line marks the 10$\times$ slowdown SLO.}
\vspace{-1em}
  \Description{p90 slowdown on a log scale against sessions completed per minute, with a dashed line at the 10 times SLO. No eviction and Continuum stay under the SLO up to their peak of 0.59 sessions per minute, and compaction at 128K up to its peak of 0.56; AgentDiet crosses it at about 0.42 and peaks at 0.46.}
  \label{fig:existing}
\end{figure}

We test three representative methods from the harness and runtime levels of \S\ref{sec:bg-approaches} that need no training: Hermes compaction~\cite{nous2026hermescompression} replaces older history with an LLM-generated summary at a 128K-token threshold (\S\ref{sec:eval-setup}); AgentDiet~\cite{xiao2026agentdiet} uses LLM reflection calls to identify and remove unnecessary history; and Continuum~\cite{li2025continuum} retains KV across tool executions for reuse by the next request, without shortening its history. Fig.~\ref{fig:existing} shows that none improves throughput under SLO over no eviction. Continuum's curve lies on no eviction's. Compaction's and AgentDiet's p90 slowdowns are above no eviction's at every load shown, and served unloaded, their p90 is 2.2$\times$ and 4.2$\times$ no eviction's. We study each method's runs and identify three causes.

\begin{figure}[t]
  \centering
  \includegraphics[width=\columnwidth]{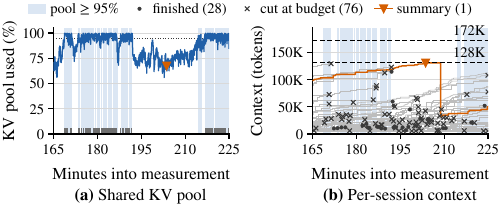}
  \vspace{-2em}
  \caption{One hour of Hermes with compaction at 128K tokens, at an arrival rate of 1.6 sessions/min. The pool is often nearly full while no session reaches the threshold, and the only summary starts after the pressure subsides.}
\vspace{-1em}
  \Description{Panel a shows shared KV occupancy, periods above 95 percent, and request preemptions. Panel b shows per-session context lengths against the 128K tuned compaction threshold and 172K default threshold. The only compaction in the selected hour begins when shared occupancy is approximately 68 percent.}
  \label{fig:batch}
\end{figure}

\noindent\textbf{Session-local decisions miss shared KV pressure.}
Compaction and AgentDiet decide from one session's logical context, while KV demand comes from all sessions. In Fig.~\ref{fig:batch}, the shared KV pool stays at least 95\% full for 20 minutes while no session reaches the compaction threshold, and the only summary begins after occupancy drops to 68\%, too late to relieve the pool. Neither sees cached-prefix residency, which decides whether a removal forces a re-prefill (\S\ref{sec:bg-stack}).

\noindent\textbf{KV retention does not reduce context demand.}
Continuum manages KV residency across concurrent sessions, but its retention policy leaves the logical context submitted in subsequent requests unchanged. Consequently, these requests continue to demand KV for the same history.

\begin{figure}[t]
  \centering
  \includegraphics[width=\columnwidth]{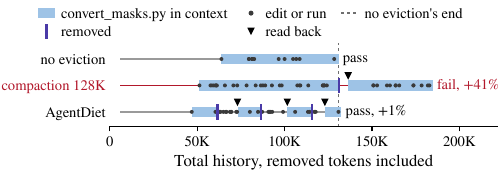}
\vspace{-2em}
  \caption{{One file of a TB2.1 task in three recorded runs; the red row failed, and percentages compare total history with no eviction's. The x-axis means the cumulative history generated during execution. Model-selected removal can drop a file that the agent then re-reads.}}
  \Description{Three trajectories show the history of convert_masks.py under no eviction, compaction at 128K, and AgentDiet. Context reduction removes the earlier copy of the script from the conversation while the agent continues editing the file, causing subsequent file reads. The 128K compaction run fails and ends with 41 percent more total history than the no-eviction run; the AgentDiet run passes with 1 percent more.}
  \label{fig:exemplar}
\end{figure}

\noindent\textbf{LLM-driven logical context reduction is costly and fallible.}
Compaction and AgentDiet shorten the history with a model's judgment. A model can weigh what the agent may need again, but its judgments on the serving GPU cost GPU time and energy, and the harness waits for each one before the session's next request. In our recordings, a compaction summary call stalls its session for a median of 91\,s. AgentDiet's reflection call, 11\,s at the median, runs before 60\% of agent requests in the median session, and served unloaded its sessions take 1.5$\times$ as long as no eviction's. The model's judgment in both methods can remove file contents from the session history that the agent still needs, forcing re-reads (Fig.~\ref{fig:exemplar}).

Lifting the limit calls for both a selection signal without extra model calls, to decide what to remove, and a global view of shared KV pressure and cached-prefix residency, to decide when. \S\ref{sec:opportunities} shows that the serving runtime can provide both.

\section{Opportunities and Challenges}
\label{sec:opportunities}

Reducing shared KV pressure through context eviction requires identifying removable logical context and deciding when to remove it. These decisions depend on agent execution history, shared runtime state, and the effects of removal on subsequent execution.

\subsection{Key Insights and Opportunities}
\label{sec:obs-evictable}
\label{sec:obs-info}
\noindent\textbf{Tool Executions Mark Reclaimable Context.}
\begin{figure}[t]
  \centering
  \includegraphics[width=\columnwidth]{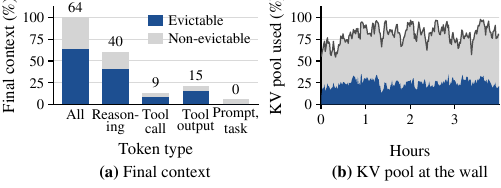}
  \vspace{-2.5em}
  \caption{{Removal candidates marked by later tool operations: (a) their share of total final-history tokens, broken down by token type, and (b) their estimated share of the KV pool in one run of no eviction at its peak throughput. Candidates are most of the context and a large share of a full pool.}}
\vspace{-1.5em}
  \Description{Panel a shows that supersession rules mark 64.3 percent of final-context tokens, including tool results and associated reasoning and tool-call history. Panel b shows that marked spans account for approximately 29 to 31 percent of the shared KV pool when occupancy reaches at least 95 percent.}
  \label{fig:reach}
\end{figure}
Tool executions reveal substantial opportunities to reduce agent context. Fig.~\ref{fig:reach}(a) shows that later tool operations mark 64\% of tokens in recorded final histories as candidates for removal, including earlier tool outputs and their associated reasoning and tool calls. These candidates also contribute significantly to memory demand under load. At no eviction's peak throughput, marked spans of in-flight requests account for an estimated 30\% of the shared KV pool's capacity when occupancy reaches at least 95\% (Fig.~\ref{fig:reach}(b)). This indicates substantial potential for reducing context demand when GPU memory is scarce.

We identify these candidates through \emph{tool-operation supersession}. Later tool operations can supersede earlier observations of the same state. For example, rereading a file provides a newer view of its contents, while a successful test rerun can supersede an earlier failed result. The earlier outputs become removal candidates, and once all tool outputs of an assistant turn are superseded, its associated tool calls and reasoning also become candidates. By matching tool operations and their targets, our rules classify 84\% of Hermes's and 83\% of OpenClaw's tool-output tokens~(see the setup details in \S\ref{sec:eval-setup}), enabling supersession detection directly from execution history without additional model inference.

\noindent\textbf{The Runtime Sees the Signal and the Pool.}
\label{sec:opp-runtime}
The superseded context identified above is visible in the requests received by the serving runtime. Each request carries structured tool calls and their outputs, allowing the runtime to identify supersession without accessing the tool sandboxes.
The runtime also manages the shared KV pool across concurrent sessions, including its memory occupancy and cached-prefix residency. This combined visibility enables context eviction decisions that account for both the memory pressure created by concurrent sessions and the cached states that a removal would invalidate.

\subsection{Challenges for Runtime Context Eviction}
\label{sec:challenges}
\label{sec:obs-prefix}
\label{sec:obs-scope}

\begin{figure}[t]
  \centering
  \includegraphics[width=\columnwidth]{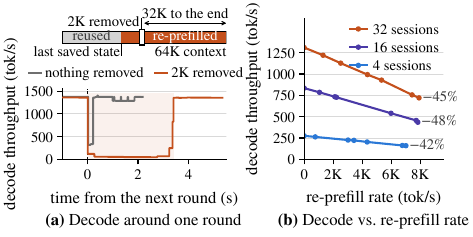}
  \vspace{-2.5em}
  \caption{{Decode throughput of Qwen on one H100 (a) around one request that removes 2K tokens from a 64K-token cached context, or nothing, with 32 sessions decoding, and (b) against the re-prefill rate. A small removal can cause a large re-prefill that stalls concurrent decoding.}}
\vspace{-1.5em}
  
  \Description{Panel a shows that removing 2K tokens from the middle of a 64K cached context causes the next round to re-prefill approximately 41K tokens instead of 1.5K. During this 3.4-second re-prefill, aggregate decoding throughput of 32 concurrent sessions drops from approximately 1360 to 53 tokens per second. Panel b shows decreasing aggregate decode throughput as sustained re-prefill rate increases.}
  \label{fig:stall}
\end{figure}
Tool-operation supersession identifies candidates for context removal, while the serving runtime observes the shared memory pressure and cached-prefix residency (\S\ref{sec:obs-evictable}). However, turning these candidates into actual removals raises two challenges: 

\noindent\textbf{Selecting context without disrupting subsequent agent actions.} Removing history changes the information available to the model in subsequent rounds. A newer tool result may supersede an earlier observation before the agent finishes the related actions, potentially forcing repeated work or disrupting task completion, as illustrated in \S\ref{sec:design-space}. Effective context eviction must therefore identify superseded information while preserving the context needed for subsequent agent actions.

\noindent\textbf{Deciding when the memory savings justify the cost of invalidating cached states.} Removing an earlier context span invalidates the cached states of the surviving suffix, introducing additional prefill that interferes with concurrent decoding. Fig.~\ref{fig:stall}(a) shows that removing just 2K tokens from a cached context triggers 41K tokens of re-prefill instead of 1.5K, sharply reducing the throughput of 32 concurrent decoding sessions. The re-prefill exceeds the 32K-token suffix because the hybrid model resumes from the last saved recurrent state before the cut. Fig.~\ref{fig:stall}(b) further quantifies this interference across different concurrency levels. As competing re-prefill increases from 0 to 8K tokens/s, aggregate decode throughput falls by 42--48\% across 4, 16, and 32 concurrent sessions. Thus, the computation introduced by context eviction can substantially reduce the serving capacity available to other sessions.

\section{System Design}
\label{sec:design}
\begin{figure}[t]
    \centering
    \resizebox{\columnwidth}{!}{\usetikzlibrary{arrows.meta,calc,decorations.pathreplacing}%
\begin{tikzpicture}[
  x=1cm, y=1cm,
  font=\fontsize{7}{8}\selectfont,
  hlayer/.style={draw=blue!45!black, semithick, fill=blue!5},
  tlayer/.style={draw=brown!60!black, semithick, fill=brown!8},
  rlayer/.style={draw=black!45, semithick, fill=gray!4},
  sess/.style={draw=blue!55!black, semithick, fill=white},
  sbox/.style={draw=brown!60!black, semithick, fill=brown!22, align=center, inner sep=1.5pt},
  box/.style={draw=black!60, semithick, fill=gray!14, align=center, inner sep=1.5pt},
  mbox/.style={draw=cyan!45!black, thick, fill=cyan!22, align=center, inner sep=1.5pt},
  ours/.style={draw=orange!80!black, very thick, fill=orange!10},
  newours/.style={box, draw=purple!70!black, very thick, fill=purple!16},
  cont/.style={draw=black!50, semithick, fill=gray!4},
  lyr/.style={font=\fontsize{8}{9.6}\selectfont\bfseries, anchor=north west, inner sep=1.5pt},
  tab/.style={font=\fontsize{8}{9.6}\selectfont\bfseries, fill=gray!4, inner sep=1.5pt, anchor=west},
  ttl/.style={font=\fontsize{8}{9.6}\selectfont\bfseries, inner sep=0pt},
  lbl/.style={inner sep=1pt, align=center},
  small/.style={font=\fontsize{7}{7.8}\selectfont},
  flow/.style={-{Stealth[length=1.5mm]}, semithick},
  ctrl/.style={-{Stealth[length=1.5mm]}, semithick, densely dashed, orange!80!black},
  fb/.style={-{Stealth[length=1.5mm]}, thick, dash dot, purple!70!black},
  async/.style={-{Stealth[length=1.5mm]}, thick, densely dotted, black!75},
  cell/.style={draw=black!55, thin, minimum size=1.8mm, inner sep=0pt, anchor=west, fill=white},
  sA/.style={cell, fill=blue!18},
  sB/.style={cell, fill=blue!50},
  sN/.style={cell, fill=blue!75!black},
  dead/.style={cell, fill=red!22, draw=red!70!black, semithick},%
  cut/.style={cell, fill=white},%
  fresh/.style={cell, fill=yellow!70!orange},
  cross/.style={red!70!black, thin},
  brc/.style={decorate, decoration={brace, amplitude=2pt}, thin},
]
\def\cw{0.18}%

\draw[hlayer] (0,4.69) rectangle (4.87,5.54);
\draw[tlayer] (6.1,4.69) rectangle (8.4,5.54);
\draw[rlayer] (0,0)    rectangle (8.4,4.37);
\node[lyr] at (0,5.54) {Agent harness};
\node[lyr] at (6.1,5.54) {Tool sandboxes};
\node[lyr] at (0,4.37) {Serving runtime};

\foreach \x/\n/\s in {0.15/1/sA, 1.55/2/sB, 3.47/N/sN} {
  \draw[sess] (\x,4.76) rectangle ++(1.25,0.46);
  \node[inner sep=0pt] at (\x+0.625,5.09) {Session $\n$};
  \foreach \i in {0,...,3} \node[\s] at ({\x+0.265+\i*\cw},4.86) {};
}
\node at (3.135,4.99) {$\cdots$};

\foreach \x/\n in {6.25/1, 6.85/2, 7.75/N} \node[sbox, minimum width=0.5cm, minimum height=0.46cm] at (\x+0.25,4.99) {$\n$};
\node at (7.57,4.99) {$\cdots$};
\draw[flow] (4.87,5.22) -- node[small, above=0.5pt, inner sep=0.5pt] {tool calls} (6.1,5.22);
\draw[flow] (6.1,4.8) -- node[small, above=0.5pt, inner sep=0.5pt] {outputs} (4.87,4.8);

\begin{scope}[shift={(0.12,3.9)}]
  \node[cell] at (0,0) {};          \node[small, anchor=west, inner sep=1pt] at (0.22,0) {context};
  \node[dead] at (0,-0.24) {};      \node[small, anchor=west, inner sep=1pt] at (0.22,-0.24) {candidate};
  \node[cut] (lx) at (1.45,0) {};
  \draw[cross] (lx.south west) -- (lx.north east) (lx.north west) -- (lx.south east);
                                    \node[small, anchor=west, inner sep=1pt] at (1.67,0) {evicted span};
  \node[fresh] at (1.45,-0.24) {};  \node[small, anchor=west, inner sep=1pt] at (1.67,-0.24) {new input};
\end{scope}

\node[box, minimum width=4.95cm, minimum height=0.3cm] (api) at (5.775,4.15) {API server};
\draw[flow] (3.8,4.3) -- (3.8,4.69);
\draw[flow] (4.05,4.69) -- (4.05,4.3);
\node[small, anchor=east, inner sep=1pt] at (3.8,4.495) {responses (reasoning, text, tool calls)};
\node[small, anchor=west, inner sep=1pt] at (4.05,4.495) {requests (session ID, tool calls, outputs)};

\draw[ours] (3.3,2.66) rectangle (6.8,3.5);
\node[ttl] at (5.05,3.33) {Session Tracker};
\foreach \r/\n/\x/\y in {1/4/4.45/3.06, 2/5/5.65/3.06, N/4/4.45/2.82} {
  \node[small, anchor=east, inner sep=1pt] at (\x-0.03,\y) {$\r$};
  \foreach \i in {1,...,\n} \node[cell] at ({\x+(\i-1)*\cw},\y) {};
}
\node[dead]  at ({4.45+1*\cw},3.06) {};
\node[dead]  at ({5.65+2*\cw},3.06) {};
\node[dead]  at ({5.65+3*\cw},3.06) {};
\node[fresh] at ({5.65+4*\cw},3.06) {};
\node[dead]  at ({4.45+3*\cw},2.82) {};

\draw[async] (3.6,4.0) -- node[small, right, inner sep=1.5pt] {after dispatch (async)} (3.6,3.5);

\draw[cont] (0.15,1.52) rectangle (7.68,2.36);
\node[tab] at (0.65,2.36) {Session Evictor};
\node[newours, minimum width=2.0cm, minimum height=0.45cm] (act) at (1.3,1.9) {Policy Actuator};
\draw[ours] (3.3,1.6) rectangle (7.55,2.2);
\node[ttl] at (5.425,2.04) {Context Filter};
\node[small, anchor=east, inner sep=1pt] at (4.97,1.77) {$2$};
\foreach \i in {0,1} \node[cell] at ({5.0+\i*\cw},1.77) {};
\foreach \i in {2,3} {
  \node[cut] (x\i) at ({5.0+\i*\cw},1.77) {};
  \draw[cross] (x\i.south west) -- (x\i.north east) (x\i.north west) -- (x\i.south east);
}
\node[fresh] at ({5.0+4*\cw},1.77) {};

\draw[ctrl] (3.3,2.8) -- (0.45,2.8) -- (0.45,2.125);
\node[small, lbl, text=red!75!black, anchor=south] at (1.87,2.82) {candidates from\\earlier requests};
\draw[ctrl] (act.east) -- node[small, above, inner sep=1pt, text=orange!70!black] {commits} (3.3,1.9);
\draw[flow] (7.1,4.0) -- (7.1,2.2);
\node[small, anchor=west, inner sep=1pt] at (7.13,3.08) {prompt};
\draw[ctrl] (6.5,2.2) -- (6.5,2.66);
\node[small, anchor=east, text=orange!70!black, inner sep=1pt] at (6.45,2.52) {mark evicted context};

\node[box, minimum width=2.05cm, minimum height=0.42cm] (kvm) at (1.175,0.98) {KV cache manager\\[-1.5pt]+ prefix cache};
\node[box, minimum width=1.3cm, minimum height=0.42cm] (sched) at (5.0,0.98) {Scheduler};
\node[mbox, minimum width=1.3cm, minimum height=0.42cm] (llm) at (7.6,0.98) {\fontsize{8}{9.6}\selectfont\bfseries Model};
\draw[flow] (sched) -- (kvm);
\draw[flow] (sched) -- node[small, above=0.5pt, inner sep=0.5pt] {batch} (llm);
\draw[flow] (8.1,1.19) -- (8.1,4.0);
\draw[flow] (5.0,1.6) -- node[small, right, inner sep=1.5pt] {shortened prompt} (sched.north);
\draw[fb, {Stealth[length=1.5mm]}-{Stealth[length=1.5mm]}] (1.3,1.19) -- (1.3,1.675);
\node[small, anchor=west, text=purple!70!black, inner sep=1pt] at (1.35,1.36) {KV usage; residency probe};

\def\px{1.85}\def\py{0.2}
\node[anchor=west, inner sep=0pt] (pool) at (0.1,\py) {shared KV pool};
\foreach \i/\s in {0/sA,1/sA,2/sA,3/sB,4/sB,5/sB,6/sN,7/sN,8/sN,9/sA,10/sB,11/sA,12/sN,13/sB,14/sA,15/sN,16/sB,17/sA,18/sN,19/sA} \node[\s] at ({\px+\i*\cw},\py) {};
\foreach \i in {20,...,25} \node[cell] at ({\px+\i*\cw},\py) {};
\draw[flow] (0.9,0.77) -- (pool.north -| 0.9,0);
\draw[flow, {Stealth[length=1.5mm]}-{Stealth[length=1.5mm]}] ({\px+26*\cw+0.04},\py) -| (llm.south);
\draw[brc] (\px,{\py+0.12}) -- node[above=1.5pt, inner sep=0pt] {running and cached} ({\px+20*\cw},{\py+0.12});
\draw[brc] ({\px+20*\cw},{\py+0.12}) -- node[above=1.5pt, inner sep=0pt] {free} ({\px+26*\cw},{\py+0.12});
\end{tikzpicture}%
} %
    \vspace{-2em}
    \caption{Architecture of \sys.}
    \vspace{-1.5em}
    \Description{The agent harness, one instance per session, sends requests, each with its session ID and full history, to the serving runtime's API server and receives responses (reasoning, text, tool calls); tool sandboxes, one per session, run the tool calls and return their outputs to the harness. On the request path, the API server passes each prompt to the Context Filter in the Session Evictor. A dotted arrow, labeled after dispatch (async), also passes the request to the Session Tracker, which keeps each session's context in its own row and marks candidates off the request path. The tracker sends candidates from earlier requests to the Policy Actuator in the Session Evictor, which reads KV cache usage from the KV cache manager, asks it through the residency probe how much of the session's prompt is cached, and sends commits to the Context Filter. The filter removes the committed spans, marks them evicted in the tracker, and passes the shortened prompt to the scheduler. The scheduler allocates KV through the KV cache manager with its prefix cache and runs the model, which reads and writes a shared GPU KV pool whose blocks belong to several sessions or are free; the model's output returns through the API server.}
    \label{fig:system-design}
\end{figure}
\label{sec:design-implications}
\sys performs context eviction within the serving runtime, using tool-operation supersession to identify removal candidates and KV cache state to decide when to remove them. Its design follows four goals derived from opportunities and challenges in \S\ref{sec:opportunities}:
\begin{enumerate}[label=\textbf{G\arabic*},leftmargin=*,nosep]
  \item \textbf{Global visibility into shared KV state}: use the runtime's view of shared KV pressure and cached-prefix residency across sessions to decide when to evict context.
  \item \textbf{Semantic selection through tool-operation supersession}: identify superseded tool outputs and their turns from execution history without model calls, keeping the context that later actions need.
  \item \textbf{Harness- and model-agnostic context eviction}: shorten only the prompt the model executes, in the runtime, without modifying the harness or the model and without training.
  \item \textbf{Bounded cache invalidation and re-prefill interference}: avoid unnecessary recomputation and bound the cached computation each eviction sacrifices relative to the context it removes.
\end{enumerate}

\subsection{Overview}
\label{sec:design-overview}

Fig.~\ref{fig:system-design} shows how \sys processes each request 
between the API server and the unmodified scheduler. Each request carries the session's full history and a session ID. 
The \emph{Session Evictor} retrieves the session's removal \emph{candidates}, and its \emph{Policy Actuator} uses shared KV pool occupancy and the candidates' cached-prefix residency to decide whether to \emph{commit} their removal or defer it (\S\ref{sec:pressure}, \S\ref{sec:commit}). The \emph{Context Filter} then removes all committed spans from the prompt before scheduling (\S\ref{sec:mech}). The harness keeps its full history, while the model processes the filtered prompt.

After dispatch, the \emph{Session Tracker} processes the request's new tool messages asynchronously 
and marks earlier outputs that later operations supersede as candidates (\S\ref{sec:tracking}).
Each commit decision therefore uses candidates from earlier requests, which keeps tracking off the critical path. 
Only a session's first request, or one whose earlier history changed, waits for tracking.

\subsection{Tracking Superseded Context}
\label{sec:tracking}
\begin{figure}[t]
  \centering
  \usetikzlibrary{arrows.meta,calc,decorations.pathreplacing}%
\begin{tikzpicture}[
  x=1cm, y=1cm,
  font=\fontsize{7}{8}\selectfont,
  sc/.style={draw=black!55, thin, minimum height=3.2mm, inner sep=0pt, text height=1.5ex, text depth=0.2ex, anchor=south west, fill=white, text=black!80},
  tsk/.style={sc, minimum width=9mm},
  rc/.style={sc, minimum width=4mm},
  fresh/.style={fill=yellow!70!orange},
  redo/.style={fill=blue!35},
  cross/.style={red!70!black, thin},
  cached/.style={line width=1.4pt, black!75},
  txt/.style={anchor=west, align=left, inner sep=0pt, text width=4.45cm},
  tag/.style={anchor=east, inner sep=0pt, font=\fontsize{7}{8}\selectfont\bfseries},
  msg/.style={draw=black!55, thin, minimum height=3.6mm, inner sep=0pt, text height=1.6ex, text depth=0.3ex, anchor=south west, fill=white},
  omsg/.style={msg, minimum width=13.6mm},
  cmsg/.style={msg, minimum width=1.6mm},
  cand/.style={fill=red!22, draw=red!70!black, semithick},
  rnd/.style={anchor=south, inner sep=0pt, text=black!60},
  sup/.style={-{Stealth[length=1.4mm]}, semithick, black!70},
  brc/.style={decorate, decoration={brace, mirror, amplitude=2pt}, thin},
]
\def\x{0.3}%
\def\tw{0.9}%
\def\cx{1.2}%
\def\xa{1.2}\def\xb{1.75}\def\xc{2.05}\def\xd{2.5}\def\xe{2.8}\def\xn{3.2}\def\xz{3.55}%
\newcommand{\strip}[1]{%
  \node[tsk] at (\x,#1) {task};
  \node[sc, cand, minimum width=5.5mm] at (\xa,#1) {1};
  \node[sc, cand, minimum width=3mm] at (\xb,#1) {2};
  \node[sc, cand, minimum width=4.5mm] at (\xc,#1) {3};
  \node[sc, minimum width=3mm] at (\xd,#1) {4};
  \node[sc, minimum width=4mm] at (\xe,#1) {5};
  \node[sc, fresh, minimum width=3.5mm] at (\xn,#1) {};
}

\def\ry{3.15}%
\node[tag] at (0.22,{\ry+0.18}) {(a)};
\foreach \k/\t/\st in {1/read app.py/cand, 2/edit app.py/cand, 3/test fails/cand, 4/edit app.py/, 5/test passes/} {
  \node[cmsg, \st] at ({\x+(\k-1)*1.62},\ry) {};
  \node[omsg, \st] (o\k) at ({\x+(\k-1)*1.62+0.16},\ry) {\t};
  \node[rnd] at ({\x+(\k-1)*1.62+0.08},{\ry+0.42}) {\k};
}
\draw[sup] ($(o2.north)+(-0.3,0)$) to[bend right=28] ($(o1.north)+(0.2,0)$);
\draw[sup] ($(o4.north)+(-0.3,0)$) to[bend right=16] ($(o2.north)+(0.3,0)$);
\draw[sup] ($(o5.south)+(-0.3,0)$) to[bend left=16] node[below=0.5pt, inner sep=0pt] {same test passes} ($(o3.south)+(0.3,0)$);

\draw[densely dashed, black!60] (\cx,1.08) -- (\cx,2.84);
\node[anchor=south, inner sep=1pt] at (\cx,2.84) {cut};

\node[tag] at (0.22,2.38) {(b)};
\strip{2.22}
\draw[cached] (\x,2.66) -- (\cx,2.66);
\node[anchor=south west, inner sep=1pt] at (\x,2.66) {cached};
\node[txt] at (3.9,2.38) {Nothing cached past the cut.\\\textbf{Commit} at any pool usage (cold commit).};

\node[tag] at (0.22,1.58) {(c)};
\strip{1.42}
\draw[cached] (\x,1.86) -- (\xn,1.86);
\node[anchor=south, inner sep=1pt] at (2.2,1.86) {cached prefix};
\draw[brc] (\cx,1.38) -- node[below=1.5pt, inner sep=0pt] {removed} (\xd,1.38);
\draw[brc] (\cx,1.12) -- node[below=1.5pt, inner sep=0pt] {invalidated} (\xn,1.12);
\node[txt] at (3.9,1.58) {Cached past the cut.\\No pool pressure: \textbf{hold} (pressure gate).\\Under pressure: \textbf{commit} within the cap.};

\def\dy{0.42}%
\node[tag] at (0.22,{\dy+0.16}) {(d)};
\node[tsk] at (\x,\dy) {task};
\foreach \k/\xx/\w in {1/\xa/5.5mm, 2/\xb/3mm, 3/\xc/4.5mm} {
  \node[sc, minimum width=\w] (d\k) at (\xx,\dy) {\k};
  \draw[cross] (d\k.south west) -- (d\k.north east) (d\k.north west) -- (d\k.south east);
}
\node[sc, redo, minimum width=3mm] at (\xd,\dy) {4};
\node[sc, redo, minimum width=4mm] at (\xe,\dy) {5};
\node[sc, fresh, minimum width=3.5mm] at (\xn,\dy) {};
\node[txt] at (3.9,{\dy+0.16}) {After the commit, the task is reused and rounds 4--5 are prefilled again (blue).};
\end{tikzpicture}%
    \vspace{-2em}
  \caption{\sys's marking and commit policy. (a) Later tool operations mark superseded outputs as removal candidates. (b)--(d) A later request, after rounds 1--3 pass the grace window. (b) A batch commits immediately when no cached tokens follow its earliest removal point. (c) Otherwise, it commits under pool pressure when the re-prefill cap is satisfied. (d) The commit reuses the prefix before the removal point and re-prefills the remaining suffix.}
  \vspace{-1.5em}
  \Description{Panel a: one session's five rounds, numbered 1 to 5 left to right, each a narrow box for the round's tool call and reasoning beside a box for its tool output: read app.py, edit app.py, test fails, edit app.py, test passes. Arrows run from the first edit back to the read, from the second edit back to the first edit, and, labeled same test passes, from the passing test back to the failed one. The first three rounds are light red with a red outline: they are candidates. Panels b to d show the same session's prompt at a later request, after rounds 1 to 3 have passed the grace window, as one cell per part, each as wide as its tokens (rounds after 5 are not drawn): the task, rounds 1 to 5, and a yellow new-input cell; rounds 1 to 3 are light red candidates, and a dashed line marks the cut before round 1. Panel b: a bar labeled cached, for the cached prefix, ends at the cut, so nothing cached is invalidated and the batch commits at any pool usage (cold commit). Panel c: the cached prefix extends through round 5; one brace, labeled removed, marks rounds 1 to 3, and a longer brace, labeled invalidated, runs from the cut to the end of the cached prefix; without pool pressure the batch holds (pressure gate), and under pressure it commits if the invalidated tokens stay within the re-prefill cap. Panel d: the committing request's prefill: the task cell is reused, rounds 1 to 3 are crossed out, and rounds 4 and 5 are shaded as prefilled again.}
  \label{fig:commit}
\end{figure}
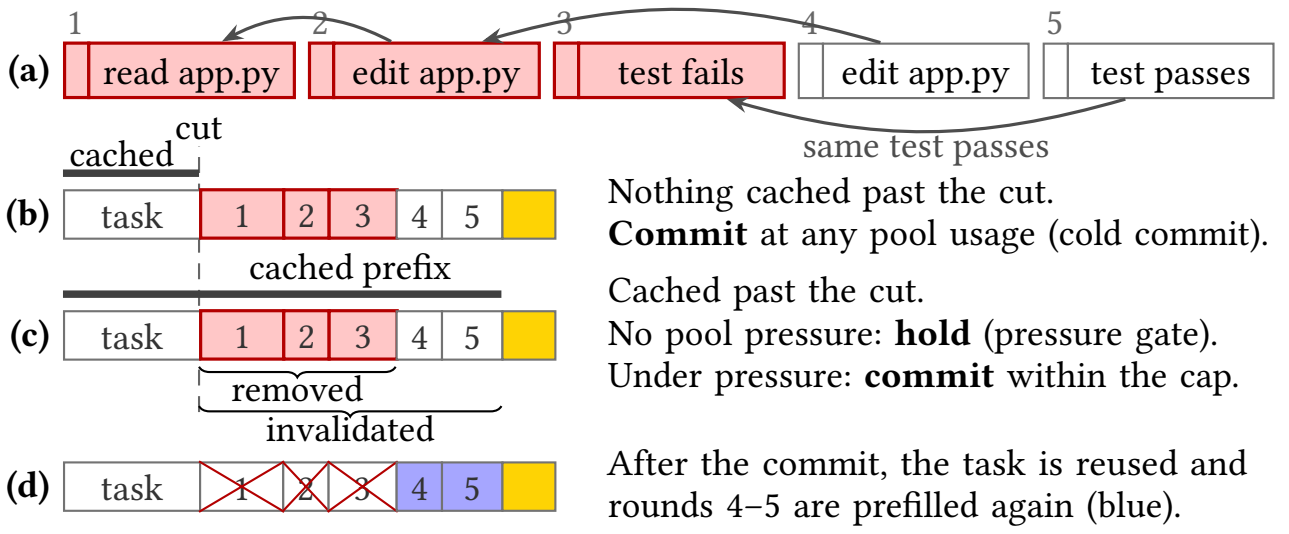

The \emph{Session Tracker} semantically identifies removal candidates from the agent's execution history without model calls (G2). For each request, it processes only the new messages and records each tool operation's type, target, and outcome.

\noindent\textbf{Observing tool operations.} 
To continue a session, the harness includes each tool call and its output in the next model request. The tracker therefore extracts tool operations directly from requests without accessing the harness or tool sandboxes.
A tool call's name and arguments identify the operation and its target. File tools name file paths, and the tracker parses shell command lines to find the programs they run and the files they read or write. Each tool output is linked to its call and records the outcome. A new harness or tool only extends the mapping from tool names and argument fields to operations. Calls to external services, such as web APIs and MCP servers, are tracked the same way.

\noindent\textbf{Supersession rules.}
The tracker matches later tool operations to earlier outputs by operation type and target (Fig.~\ref{fig:commit}(a)). A later read or edit of a file supersedes earlier observations of it. Rerunning a normalized shell command supersedes its previous output, and a later run or build supersedes earlier run or build outputs whose target files overlap. Failed outputs are not superseded by unrelated operations, and a test result is superseded only when a rerun of the same test passes. Once all tool outputs of a round are superseded, the round's tool calls and reasoning also become candidates. An unmapped tool's output is superseded only by a repeat of the same call or a later run over a file the call names; otherwise it stays. The system prompt and task description are always retained.

\noindent\textbf{Grace window.}
Candidates become \emph{mature}, and eligible for commit, only after a configurable grace window, 2 rounds by default (\S\ref{sec:eval-knobs-selection}).
The window keeps recently superseded information while the agent completes related operations, such as verifying an earlier edit. We evaluate the sensitivity of \sys to it in \S\ref{sec:eval-knobs-selection}.

\noindent\textbf{Locating spans.} 
To remove a span, \sys must know where each message lies in the token prompt that the chat template renders. The tracker renders the conversation a second time with the new message's content blanked and takes the window where the two renders differ as its span, which leaves the template's tool-response markers intact. Because tracking is asynchronous, spans come from an earlier prompt. The filter removes a span only while the current prompt still matches that prompt up to the span's end. Otherwise, the request waits for tracking.

\subsection{Eviction Timing from Global KV State}
\label{sec:pressure}

\sys uses the runtime's global view of shared KV state (G1) to decide when mature candidates are removed. Pool pressure determines when reducing context demand is beneficial, and cached-prefix residency determines whether a removal sacrifices reusable computation. The pressure gate and cold commits use these signals to avoid unnecessary re-prefill (G4).

\noindent\textbf{Pressure gate.}
When a removal would invalidate cached states, \sys defers it until the shared KV pool is under pressure. With memory to spare, the invalidation adds computation without relieving capacity. Under pressure, removing superseded context reduces the KV demand of later requests, which admits more concurrent sessions and avoids preemptions.
Pressure is the fraction of KV blocks held by running requests, with a configurable threshold of 80\% by default. We evaluate \sys's sensitivity to it in \S\ref{sec:eval-knobs-pressure}.
Cached blocks of idle sessions are excluded because the allocator can reclaim them on demand. This global pressure signal coordinates eviction across sessions, while each session applies eligible removals at its next request.

\noindent\textbf{Cold commits.}
When the cached prefix ends before the earliest removal point, the next request recomputes the uncached suffix regardless, e.g., after cached blocks are reclaimed during tool execution. Committing cuts this recomputation without sacrificing cache reuse, so \sys makes such cold commits regardless of pressure.

\subsection{Cache-Aware Commit Policy}
\label{sec:commit}

To balance context reduction against lost cache reuse (G4), \sys groups a session's matured candidates into one \emph{batch} (Fig.~\ref{fig:commit}(c)). The batch's \emph{gain} is the number of prompt tokens it removes. Since each prefix-cache block depends on all preceding tokens, the batch's \emph{cost}, the cached tokens whose reuse it invalidates, equals the number of cached tokens after the earliest removal point, or zero if the cached prefix ends at or before that point. Candidates after the earliest removal point therefore add gain without adding cost.

\noindent\textbf{Residency probe.} The cached-prefix length changes at each engine step as sessions run and the allocator reclaims idle blocks. The API server cannot compute it because the engine chains block hashes with a per-process seed, so \sys adds a read-only \emph{residency probe} to the engine's KV cache manager. The probe hashes a session's prompt block by block, as the scheduler does, and looks the hashes up in the GPU prefix cache. It neither allocates blocks nor modifies cache state, and the Policy Actuator invokes it only when a session has a matured batch.

\noindent\textbf{Commit rule.} 
The \emph{Policy Actuator} commits a batch whose cost is zero regardless of pressure (a cold commit, Fig.~\ref{fig:commit}(b)).
Otherwise, it commits only when pool pressure exceeds the threshold and the cost is at most the gain times a configurable \emph{re-prefill cap}, 6 by default. We evaluate sensitivity to the cap in \S\ref{sec:eval-knobs-cap}. For hybrid models, vLLM saves the recurrent states of linear-attention layers only at block-aligned engine-step boundaries. Re-prefill therefore starts from the last saved state before the cut, so the cost is a lower bound on the recomputation (Fig.~\ref{fig:stall}). The cold-commit condition remains exact because a cached prefix ending before the cut also ends at a saved recurrent state. Context-window commits and tracker failures (\S\ref{sec:mech}) bypass the cap. A batch that fails these conditions stays pending until the session's next request.

\subsection{Eviction Mechanism}
\label{sec:mech}

To keep eviction harness- and model-agnostic (G3), the runtime removes committed spans from the model's input only, keeping the prompt structure valid and the prefix cache consistent.

\noindent\textbf{Prompt filtering.}
The \emph{Context Filter} removes committed spans from the token prompt after chat template rendering and before scheduling. For tool outputs, it removes their content while retaining the template's tool-response markers; committed assistant turns are removed entirely. The filter also excludes previously committed spans from every subsequent request, preventing their reintroduction from breaking prefix-cache reuse. The harness keeps and resends its full history, while the model receives the shortened prompt. The harness only adds a session ID to each request.

\noindent\textbf{Cache interaction.}
The shortened prompt follows the engine's normal prefix-caching mechanism. The committing request reuses the cached prefix before the earliest removal and re-prefills the rest (Fig.~\ref{fig:commit}(d)). Subsequent requests extend the shortened prompt and reuse its cached states. 
Context removal does not immediately release GPU memory. The removed spans' cached blocks are no longer referenced and are reclaimed through normal cache eviction. Shorter prompts reduce the KV demand of subsequent rounds and the KV entries read by full-attention layers for each generated token, lowering decode memory traffic. 

\noindent\textbf{Request handling.}
\sys checks the context window after filtering, so committed removals count before the engine applies the model's context limit. If the prompt and output budget would still exceed the limit, \sys commits the session's matured candidates regardless of pressure and the re-prefill cap. If the residency probe times out, the batch stays pending and the request proceeds with earlier commits. The probe's reading can be stale by the time the scheduler admits the request, but only in one direction. A session has at most one request in flight, so other requests can only reclaim its idle blocks. A stale reading thus only overstates the cost, so a pressure-driven commit stays within the re-prefill cap and a cold commit stays cold. If candidate tracking fails, the request runs unmodified with its full history, re-prefilling from the earliest restored span. It is rejected only if it exceeds the context window.

\subsection{Implementation}
\label{sec:implementation}

We implement \sys in vLLM~0.26~\cite{kwon2023efficient} with about 7K lines of Python. The Session Tracker and Session Evictor hook into vLLM's OpenAI-compatible API server between chat-template rendering and scheduling. The tracker runs 
on a worker thread
and parses shell commands with Python's \texttt{shlex}. The prototype has supersession rules only for sandbox and file tools; external and MCP services are out of scope. The residency probe adds fewer than 100 lines to the engine core, which answers probes between execution steps. It is the only engine-core change, and the scheduler, KV cache allocator, and model execution are unchanged. Pool usage comes from vLLM's existing stat logger, which records the scheduler's usage at each step. Each request carries its session ID in an \texttt{agent\_tracker} field, which Hermes and OpenClaw forward through configuration. Per-session tracking state lives in host memory in a single API server process.

\section{Evaluation}
\label{sec:evaluation}

We evaluate \sys across eight configurations spanning two models, two agent harnesses, and two benchmarks. We first compare its throughput under SLO against existing context-management and KV-management baselines, including observation masking (\S\ref{sec:eval-e2e}). We then evaluate its energy efficiency and task accuracy (\S\ref{sec:eval-energy}, \S\ref{sec:eval-quality}). Ablation and parameter-sensitivity experiments examine the contributions and robustness of its design (\S\ref{sec:eval-ablation}, \S\ref{sec:eval-params}). Finally, we evaluate \sys with larger HBM capacity and KV offloading, and measure its request-path overhead (\S\ref{sec:eval-offload}, \S\ref{sec:eval-overhead}).

\subsection{Experimental Setup}
\label{sec:eval-setup}

We evaluate \sys on two long-horizon agent benchmarks, two open-source agent harnesses, and one dense and one mixture-of-experts (MoE) model deployed on two GPU generations under varying load, and report throughput, latency slowdown, energy efficiency, and task accuracy.

\noindent\textbf{Workloads and platforms.} We evaluate \sys on two long-horizon agent benchmarks, Terminal-Bench~2.1 (TB2.1)~\cite{merrill2026terminal} and SWE-Atlas-QnA~(SWE-Atlas)~\cite{raghavendra2026sweatlas}, using their default task verifiers and rubrics, respectively. Both benchmarks comprise long-horizon tasks whose sessions reach 100K to 228K tokens of context at p90. We run two open-source agent harnesses, Hermes~v2026.10.11~\cite{teknium2026hermesreasoning} and OpenClaw~v2026.7.1~\cite{openclaw2026compaction}, on both benchmarks. 
The experiments use a dense model, Qwen3.8-27B~\cite{qwen2026qwen38}, on an H200 GPU and an MoE model, Qwen3.6-35B-A3B~\cite{qwen2026qwen36}, on an H100 GPU. Both models use FP8 weights and vLLM~0.26~\cite{kwon2023efficient}, with KV pools equivalent to six to seven full 262K-token contexts. They use hybrid linear and full attention (\S\ref{sec:commit}).

\noindent\textbf{Baselines.} 
We compare \sys against no eviction, production harnesses' compaction, tuned harness compaction, observation masking, and academic works: LLM-driven history editing and runtime KV retention. These are the harness- and runtime-level methods (\S\ref{sec:bg-approaches}) that run on original Hermes and OpenClaw, require no model training, and support hybrid-attention models we serve:

\begin{itemize}[leftmargin=0cm,itemindent=.2cm,labelwidth=\itemindent,labelsep=0cm,align=left]
\setlength\itemsep{0pt}
\item \emph{No eviction} retains the full history except when the harness handles context overflow. 
\item \emph{Default compaction} follows each harness's shipped trigger. Hermes compacts when a prompt reaches 75\% of the input capacity after reserving output tokens (172K tokens)~\cite{nous2026hermescompressor}, while OpenClaw compacts only upon context overflow. These triggers depend on context length rather than KV pressure and are reached by 16\% of no-eviction sessions in Hermes and 7\% in OpenClaw, on average.
\item \emph{Tuned compaction} lowers the default trigger to compact more sessions. 
We consider thresholds reached by more than 16\% of no-eviction sessions, the higher of the two defaults' mean reach, but by fewer than 40\%, as more aggressive compaction reduces accuracy~\cite{kang2026acon}. For each configuration, we select the threshold that completes the most tasks within their task budgets in unloaded recordings.
\item \emph{Observation masking}~\cite{lindenbauer2025complexity} replaces every tool output except the 10 most recent with a short placeholder, without model calls. Hermes and OpenClaw ship variants~\cite{nous2026hermescompression,openclaw2026pruning}. We implement it in vLLM's request path, so both harnesses run unmodified.
\item \emph{AgentDiet}~\cite{xiao2026agentdiet} uses a reflection call to the agent's own model on the serving GPU to reduce history, and is ported from Trae Agent~\cite{gao2025trae} to both harnesses.
\item \emph{Continuum}~\cite{li2025continuum} uses the authors' published implementation, whose default retains a completed request's KV cache for 2\,s when the tool's mean execution time is shorter than 2\,s.
\end{itemize}
\sys uses the same default parameters across all configurations: a pressure threshold of 80\%, a re-prefill cap of 6, and a grace window of 2 rounds (\S\ref{sec:design}). We evaluate \sys's sensitivity to them in \S\ref{sec:eval-params}.

\noindent\textbf{Load generation.} We use trace replay to make the workload reproducible across load sweeps and repeated runs. Fixing agent trajectories, output lengths, and tool delays prevents run-to-run variation in generated work from confounding measurements of serving performance; scheduling, caching, and \sys's eviction decisions still respond dynamically to load. We first record each task's session, including its prompts and tool execution times, for each harness, model, and baseline that changes the conversation, and then replay these recordings. Compaction and AgentDiet replay their own recordings, which capture their context modifications and additional model calls; these do not depend on load, so we replay each request by its prompt and output lengths, keeping the prefix it shares with the one before. Since \sys makes eviction decisions based on shared KV pressure, its throughput experiments replay no-eviction recordings with their full prompts, from whose content it selects what to evict, with eviction applied dynamically at runtime. No eviction, Continuum, and observation masking, which masks outputs at runtime, replay the same recordings, so all four start from the same recorded request sequence at each load. Replay preserves the recorded requests, output lengths, and inter-request gaps, including tool execution time. Sessions arrive according to a Poisson process~\cite{luo2026agentix,li2025continuum}, with identical arrival schedules across policies at each load level. A session's \emph{task budget} is the benchmark-defined task timeout~\cite{tbench2025timeouts}. Each run includes a 2\,h warmup to allow the KV pool to fill and measurements to stabilize.\footnote{Replay fixes trajectories and therefore excludes behavioral changes induced by eviction. We assess this approximation with live probes across all eight configurations and fully live execution for Qwen with Hermes on TB2.1 at three load levels. In the latter experiment, throughput is 9--13\% lower than replay for \sys and 2--5\% lower for no eviction, but \sys retains a 13\% advantage in throughput under the 10$\times$ p90 slowdown SLO.}

\begin{figure*}[t]
  \centering
  \includegraphics[width=\textwidth]{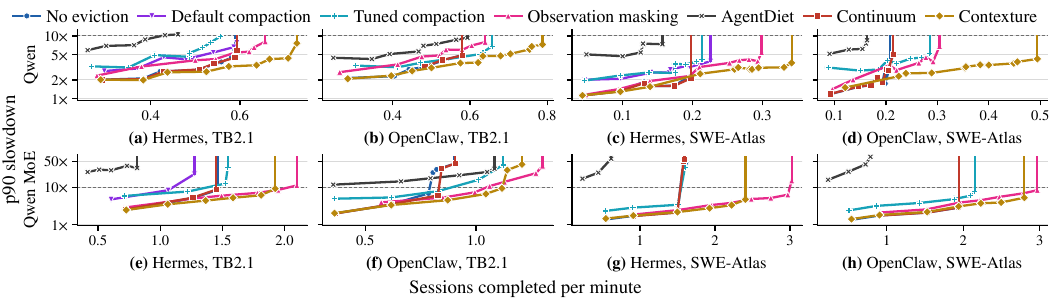}
  \vspace{-2em}
  \caption{Each policy's p90 slowdown against throughput, on a log scale, with Qwen (top) and Qwen MoE (bottom) on each workload; the dashed line is the 10$\times$ SLO. \sys sustains the highest throughput within the SLO on every workload, except three with Qwen MoE, where observation masking leads at the cost of dropping accuracy by 9--28 percentage points~(\S\ref{sec:eval-quality}).
  }
  \vspace{-1em}
  \Description{Two rows of four panels, p90 slowdown on a log scale against sessions completed per minute, with a dashed line at the 10 times SLO: Qwen on an H200 in the top row, panels (a) to (d), and Qwen MoE on an H100 in the bottom row, panels (e) to (h); in each row, Hermes and OpenClaw on Terminal-Bench 2.1, then Hermes and OpenClaw on SWE-Atlas-QnA. Qwen. Peak throughput in sessions completed per minute. Hermes, Terminal-Bench: AgentDiet 0.46, tuned compaction (128K) 0.56, no eviction, Continuum and default compaction 0.59, Contexture 0.73. OpenClaw, Terminal-Bench: no eviction and Continuum 0.58, AgentDiet 0.59, tuned compaction (96K) 0.66, Contexture 0.79. Hermes, SWE-Atlas: AgentDiet 0.16, no eviction and Continuum 0.20, tuned compaction (156K) 0.21, default compaction 0.23, Contexture 0.34. OpenClaw, SWE-Atlas: AgentDiet 0.16, no eviction 0.21, Continuum 0.21, tuned compaction (128K) 0.29, Contexture 0.49. AgentDiet runs above the other policies on every panel. Qwen MoE. Peak throughput in sessions completed per minute. Hermes, Terminal-Bench: AgentDiet 0.82, default compaction 1.28, Continuum 1.45, no eviction 1.46, tuned compaction (128K) 1.54, Contexture 1.93. OpenClaw, Terminal-Bench: no eviction 0.90, Continuum 0.91, AgentDiet 1.08, tuned compaction (128K) 1.12, Contexture 1.21. Hermes, SWE-Atlas: AgentDiet 0.62, Continuum 1.59, tuned compaction (104K) 1.60, no eviction 1.61, Contexture 2.41. OpenClaw, SWE-Atlas: AgentDiet 0.79, no eviction and Continuum 1.94, tuned compaction (144K) 2.15, Contexture 2.80. Continuum's curve lies on no eviction's in every panel, and AgentDiet runs above the other policies on every panel. Observation masking peaks at 0.66, 0.64, 0.30 and 0.30 sessions per minute with Qwen, and at 2.10, 1.30, 3.02 and 2.97 with Qwen MoE, where it crosses the SLO at 2.03, 1.05, 3.00 and 2.97, above Contexture in panels (e), (g) and (h).}
  \label{fig:eval-throughput}
\end{figure*}

\noindent\textbf{Metrics.} We report three metrics.
\emph{1. Throughput under SLO} measures the number of sessions completed within their task budgets per minute. We report the throughput at which p90 session slowdown reaches 10$\times$. Following Splitwise~\cite{patel2024splitwise}, we measure slowdown relative to the same task's no-eviction recording served unloaded. We choose the 10$\times$ SLO following prior serving systems~\cite{qin2024mooncake,agrawal2024taming,peng2026hexagent}.
\emph{2. GPU energy per completed session} is the total GPU energy divided by the number of completed sessions during the measurement window. We sample the GPU's cumulative energy counter through NVML~\cite{nvidia2026nvml,chung2025ml,mlenergy2026zeus} every second. We also report mean GPU power, calculated by dividing energy by the measurement-window duration.
\emph{3. Task accuracy} is the fraction of probe tasks that pass the benchmark's default evaluation procedure. Replay fixes every model output, so we run live agent sessions, called probes, alongside the load generator to measure accuracy under eviction.

\subsection{End-to-End Results}
\label{sec:eval-perf}

\subsubsection{Throughput and Latency Slowdown}
\label{sec:eval-e2e}
We evaluate whether \sys increases throughput under SLO across two models, two harnesses, and two benchmarks. Fig.~\ref{fig:eval-throughput} compares each policy's p90 slowdown against throughput. At light load, \sys keeps p90 slowdown within 10\% of the no-eviction value by committing only removals that invalidate no cached KV (\S\ref{sec:pressure}). Under the same conditions, compaction incurs 1.4--2.5$\times$ the p90 slowdown of no eviction, and AgentDiet incurs at least 2$\times$; observation masking incurs up to 1.24$\times$, as each request moves its window and breaks the cached prefix. At the throughput where the best baseline other than observation masking reaches the SLO in each configuration, \sys achieves 30--78\% lower p90 slowdown, and 32--61\% lower where masking reaches it, in the five configurations where \sys sustains that throughput. At the 10$\times$ p90 slowdown limit, \sys completes 20--73\% more sessions per minute than that baseline in each of the eight configurations.

Fig.~\ref{fig:eval-throughput}(a) illustrates the performance differences on TB2.1 when Hermes uses Qwen. Without eviction, the shared KV pool fills and request preemptions prevent throughput from increasing, while p90 slowdown rises sharply. Continuum follows nearly the same curve because retaining a turn's KV for 2\,s does not reduce the context carried into subsequent requests. Compaction starts with more than twice the unloaded p90 slowdown of no eviction, and its context-length trigger is reached too late to relieve KV pressure. AgentDiet exhibits a different limitation. Its KV pool never fills in this workload, but reflection calls make unloaded sessions 1.5$\times$ as long as those under no eviction, causing it to reach the SLO at 28\% lower throughput. \sys delays KV pool saturation by committing deferred removals once shared memory pressure builds, achieving 23\% higher throughput under SLO than the best of them.

The same performance trends appear across the eight configurations. The better of default and tuned compaction improves throughput under SLO over no eviction by 11--37\% in five configurations, but ties or trails it in the other three. Default compaction matches no eviction in OpenClaw because both use the same overflow handling. It also matches no eviction for Hermes with Qwen MoE on SWE-Atlas because the compaction threshold is never reached. Continuum remains within 6\% of no eviction across all configurations. AgentDiet trails no eviction by 21--28\% on three Qwen workloads and leads it by only 2\% on the fourth. Across the four Qwen MoE configurations, AgentDiet's median unloaded session duration is 1.7--6$\times$ the corresponding no-eviction duration, causing it to miss the SLO even at the lightest load. Observation masking improves throughput under SLO over no eviction by 10--95\%. \sys outperforms it by 7--63\% in five configurations. In the remaining three, all using Qwen MoE, observation masking achieves 6--25\% higher throughput at the cost of 9--28 percentage points lower task accuracy (\S\ref{sec:eval-quality}). These gains reflect two effects of context reduction. It reduces per-session KV demand, delaying pool saturation under memory pressure, and reduces KV-cache reads during decoding, easing memory-bandwidth pressure and lowering time per output token (\S\ref{sec:eval-abl-tail} and \S\ref{sec:eval-gpu}).

\begin{figure}[t]
  \centering
  \includegraphics[width=\columnwidth]{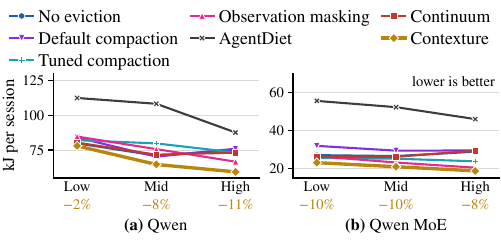}
  \vspace{-2em}
  \caption{GPU energy per completed session for each policy at Low, Mid, and High loads with Hermes on TB2.1. Numbers below \sys's points show its energy reduction relative to the lowest-energy baseline at each load.}
  \Description{Two panels of lines, (a) Qwen and (b) Qwen MoE, of GPU energy per completed session in kilojoules at Low, Mid and High load for no eviction, default compaction, tuned compaction, observation masking, AgentDiet, Continuum and Contexture; each y axis starts just under its lowest line. Qwen: AgentDiet 112, 108 and 88; observation masking 85, 76 and 67; the other baselines between 70 and 84; Contexture 78, 65 and 59, with 2, 8 and 11 percent below the baseline that spends the least printed under the loads. Qwen MoE: AgentDiet 56, 52 and 46; observation masking 26, 23 and 20; the other baselines between 24 and 32; Contexture 23, 21 and 19, with 10, 10 and 8 percent below the baseline that spends the least printed under the loads.}
  \label{fig:eval-energy-tb}
\end{figure}
\subsubsection{Energy Efficiency}
\label{sec:eval-energy}

We examine whether the throughput gains from \sys translate into lower GPU energy per completed session, accounting for the re-prefill introduced by context eviction. Fig.~\ref{fig:eval-energy-tb} reports energy per completed session for Hermes on TB2.1 at three offered loads, Low, Mid, and High, on both models. For each model, Low is the load at which no eviction reaches 90\% of its throughput under SLO, while High is the load at which \sys reaches 90\% of its throughput under SLO. Mid is the midpoint between these two offered loads.

\sys consumes the least energy per completed session at all three loads on both models. Compared with the lowest-energy baseline at each load, \sys reduces energy per completed session by 2\%, 8\%, and 11\% with Qwen, and by 10\%, 10\%, and 8\% with Qwen MoE, respectively. 
These results show that the throughput gains from \sys translate into lower GPU energy per completed session, despite the additional re-prefill caused by eviction.

Mean GPU power varies across policies by at most 2\% at each load with Qwen. With Qwen MoE, the maximum difference reaches 17\% at High, and \sys draws the least power at Mid and High. With Qwen, \sys completes only 2--11\% more sessions than the baselines other than AgentDiet at Low, but its throughput advantage grows to 12--27\% at High. This corresponds to an increase in energy savings from 2\% to 11\%. From Low to High, \sys's energy per completed session falls by 24\% with Qwen and 19\% with Qwen MoE. Compaction reduces energy per session by at most 11\%, while no eviction and Continuum reduce it by at most 9\% with Qwen and increase it with Qwen MoE. 
Despite the additional re-prefill caused by eviction, \sys performs the least prefill per completed session at every load. At High, its prefill is half that of no eviction with Qwen and one-quarter with Qwen MoE. AgentDiet consumes 1.4--2.5$\times$ as much energy per completed session as \sys. Its reflection calls add prefill and lengthen sessions (\S\ref{sec:eval-e2e}), contributing to its high energy consumption.

\begin{figure}[t]
  \centering
  \includegraphics[width=\columnwidth]{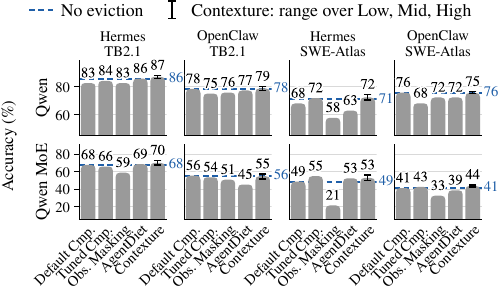}
  \vspace{-2em}
  \caption{Each policy's accuracy with Qwen and Qwen MoE on each workload. The baselines are served unloaded; \sys's bar is its mean over Low, Mid and High load and its whisker spans its lowest to its highest accuracy among the three loads. \sys under load averages 1.4 percentage points above no eviction, with a worst-case loss of 3.7.
  \vspace{-1em}
  }
  \Description{Eight tiles of accuracy bars in one column, a row per model (Qwen, Qwen MoE) and a column per workload: Hermes and OpenClaw on Terminal-Bench 2.1, then Hermes and OpenClaw on SWE-Atlas-QnA. The Qwen row's axis runs from 45\%, the Qwen MoE row's from 5\%. Each tile has a dashed line at no eviction served unloaded and five gray bars, named at an angle under the bottom row as Default Cmp., Tuned Cmp., Obs. Masking, AgentDiet and Contexture: default compaction, tuned compaction, observation masking and AgentDiet, served unloaded, and Contexture, its mean over Low, Mid and High load with a whisker from its lowest to its highest accuracy among the three. In OpenClaw, and with Qwen MoE in Hermes on SWE-Atlas-QnA, default compaction is no eviction, so its bar is at no eviction's accuracy. Qwen, Hermes, Terminal-Bench: no eviction 86\%; default compaction 83, tuned compaction 84, observation masking 83, AgentDiet 86; Contexture 87 (86 to 88). OpenClaw, Terminal-Bench: no eviction 78\%; default compaction 78, tuned compaction 75, observation masking 76, AgentDiet 77; Contexture 79 (77 to 80). Hermes, SWE-Atlas: no eviction 71\%; default compaction 68, tuned compaction 72, observation masking 58, AgentDiet 63; Contexture 72 (71 to 74). OpenClaw, SWE-Atlas: no eviction 76\%; default compaction 76, tuned compaction 68, observation masking 72, AgentDiet 72; Contexture 75 (75 to 76). Qwen MoE, Hermes, Terminal-Bench: no eviction 68\%; default compaction 68, tuned compaction 66, observation masking 59, AgentDiet 69; Contexture 70 (68 to 73). OpenClaw, Terminal-Bench: no eviction 56\%; default compaction 56, tuned compaction 54, observation masking 51, AgentDiet 45; Contexture 55 (52 to 57). Hermes, SWE-Atlas: no eviction 49\%; default compaction 49, tuned compaction 55, observation masking 21, AgentDiet 53; Contexture 53 (50 to 56). OpenClaw, SWE-Atlas: no eviction 41\%; default compaction 41, tuned compaction 43, observation masking 33, AgentDiet 39; Contexture 44 (43 to 45).}
  \label{fig:eval-accuracy}
\end{figure}

\subsubsection{Task Accuracy}
\label{sec:eval-quality}
We evaluate whether \sys preserves task accuracy when context eviction occurs under memory pressure. We run live probes on tasks solved in at least one recording and allow each task up to 6h to complete. Fig.~\ref{fig:eval-accuracy} compares the accuracy of different policies. The baselines are measured unloaded because their context-management decisions do not depend on offered load. For \sys, we report the mean and range across Low, Mid, and High loads. Across eight configurations and three loads, \sys averages 1.4 percentage points above unloaded no eviction and matches or exceeds it in most cases. Its average accuracy is 2 points higher at Low and 1 point higher at High, where it removes the most context. The largest accuracy loss occurs with OpenClaw on TB2.1 at Mid, reaching 1 point with Qwen at High and 3.7 points with Qwen MoE at Mid. With Qwen MoE, \sys matches no eviction again at High. Overall, \sys maintains task accuracy within 3.7 percentage points of unloaded no eviction across the evaluated configurations and loads.

Tuned compaction and AgentDiet exhibit larger accuracy deviations from no eviction, even when served unloaded. Tuned compaction ranges from 8 points below to 6 points above no eviction, while AgentDiet ranges from 11 points below to 4 points above. Both use model-generated context reduction, which can remove information needed by subsequent agent actions.
Observation masking also loses 3--28 percentage points across the eight configurations because its fixed window removes older tool outputs regardless of whether subsequent actions still need them. 
\sys identifies removable history through subsequent tool operations and retains candidates for a grace window before committing their removal. This preserves recently superseded context while allowing older redundant history to be removed under memory pressure.

\subsection{Ablation Study}
\label{sec:eval-ablation}
\label{sec:eval-reprefill}
\label{sec:eval-timing}

\begin{figure}[t]
  \centering
  \includegraphics[width=\columnwidth]{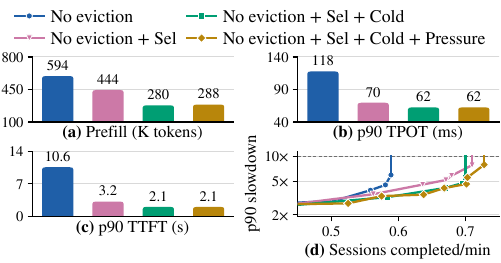}
  \vspace{-2.3em}
  \caption{Ablation of \sys with Qwen and Hermes on TB2.1. Panels (a), (b), and (c) show prefill per completed session, p90 time per output token (TPOT), and p90 TTFT at High load, pooled across three traces. Panel (d) shows p90 slowdown versus throughput, with the 10$\times$ SLO marked by a dashed line. Selection raises peak throughput, cold commits cut prefill and TTFT, and the pressure gate further increases peak throughput.}
  \vspace{-1.8em}
  \Description{Four panels, the legend above in two columns: no eviction, no eviction plus selection, plus cold commits, and plus the pressure gate, which is Contexture. (a) Prefill computed per completed session in thousands of tokens, axis from 100: 594, 444, 280 and 288. (b) p90 time per output token (TPOT) in milliseconds, axis from 40: 118, 70, 62 and 62. (c) p90 time to first token (TTFT) in seconds, axis from 0: 10.6, 3.2, 2.1 and 2.1. (d) p90 slowdown on a log scale from 2 to 10 times against sessions completed per minute from 0.45 to 0.74, with a dashed line at the 10 times SLO; peaks, equal to throughput under SLO: no eviction 0.59, plus selection 0.71, plus cold commits 0.70, plus the pressure gate 0.73; at a p90 slowdown of 5 times: 0.58, 0.66, 0.70 and 0.70.}
  \label{fig:eval-abl-compute}
  \label{fig:eval-abl}
\end{figure}
We conduct an incremental ablation to determine which mechanism raises throughput and what the cache-aware commits add at that throughput, using Qwen with Hermes on TB2.1. All configurations replay identical session arrivals. Starting from no eviction, we add one mechanism at a time. \emph{No eviction + Sel} adds candidate selection and commits superseded context as soon as it matures (\S\ref{sec:tracking}). This configuration includes file-aware supersession~\cite{cline2026contextmanager,sweagent2026historyprocessors,liu2026when} and also selects superseded command, build, and test outputs, without consulting runtime cache state. \emph{No eviction + Sel + Cold} adds cold commits, which defer removal until it can proceed without invalidating cached KV, as determined by the residency probe (\S\ref{sec:pressure}, \S\ref{sec:commit}). Finally, \emph{No eviction + Sel + Cold + Pressure} adds the pressure gate, allowing removals that invalidate cached KV when the shared pool is under pressure, subject to the re-prefill cap (\S\ref{sec:pressure}, \S\ref{sec:commit}). This final configuration corresponds to \sys. We examine the effects of each mechanism on prefill work and decoding, followed by throughput under SLO and GPU energy per completed session.

\subsubsection{Prefilling and Decoding}
\label{sec:eval-abl-prefill}
\label{sec:eval-abl-tail}

At High load, the KV pool under no eviction is full, causing the engine to preempt requests and recompute their prompts when they resume. 
 Removing superseded context allows more sessions to fit in the shared KV pool, eliminating nearly all preemptions. The shorter contexts also reduce prefill per completed session by 25\% and p90 TPOT by 41\% (Fig.~\ref{fig:eval-abl-compute}(a, b)). Fewer preemptions also cut p90 TTFT, the time a request spends queued and in prefill, by 70\% (Fig.~\ref{fig:eval-abl-compute}(c)). However, committing candidates as soon as they mature invalidates the cached prefix following each removal. This re-prefill delays the request that carries it and interferes with concurrent decoding, as shown in Fig.~\ref{fig:stall}. Cold commits avoid this cost by shortening prompts only when the affected suffix would need to be recomputed anyway (\S\ref{sec:commit}). They further reduce prefill tokens per completed session by 37\%, and p90 TPOT by 10\%. At the lightest load, where the pool never fills, selection alone doubles p90 TTFT over no eviction, while cold commits leave it unchanged. Thus, selection relieves memory pressure, while cold commits remove the re-prefill that selection adds.

\subsubsection{Throughput}
\label{sec:eval-abl-speed}
\label{sec:eval-abl-throughput}
Fig.~\ref{fig:eval-abl}(d) shows how the reductions in KV demand and re-prefill translate into higher throughput. Selection increases throughput under SLO by 20\%. 
Cold commits leave peak throughput approximately unchanged
but raise it below the peak: at a p90 slowdown of 5$\times$, they increase throughput by 6\% over selection alone. The pressure gate commits these pending removals once the threshold is crossed. Compared with cold commits alone, it increases the share of prompt tokens removed from 21\% to 25\% and reduces request preemptions by 25\%. Fewer preemptions reduce repeated prompt recomputation, slowing the increase in p90 slowdown and raising peak throughput by a further 4\% on each of the three traces.
Overall, \sys achieves 24\% higher throughput under SLO than no eviction. 
Mean GPU power differs by only about 1\% across the four configurations, so energy per completed session closely follows throughput. Selection reduces energy per session by 18\%, and cold commits provide a further 3\% reduction. The pressure gate maintains this saving at High load. Selection accounts for most of the throughput improvement, while cold commits and pressure gating reduce the associated execution overhead, cutting prefill by about one-third and p90 TTFT by about one-third relative to selection alone.

\begin{figure}[t]
  \centering
  \includegraphics[width=\columnwidth]{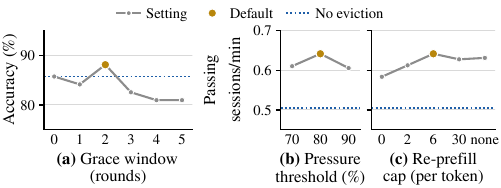}
  \vspace{-2em}
  \caption{\sys with one parameter varied at a time, compared with no eviction (dotted line; accuracy measured unloaded). (a) Task accuracy at High across grace windows. (b,c) Passing sessions per minute at High across pressure thresholds and re-prefill caps, respectively. Passing sessions per minute is calculated as throughput under SLO multiplied by task accuracy. Gold markers indicate the default settings.}
  \vspace{-1.6em}
  \Description{Three line charts, the default in gold and no eviction as a dotted line. (a) Accuracy in percent at grace windows of 0 to 5 rounds: 86, 84, 88 (the default), 83, 81 and 81; no eviction 86. (b) Passing sessions per minute at pressure thresholds of 70, 80 (the default) and 90 percent: 0.61, 0.64 and 0.61; no eviction 0.50. (c) Passing sessions per minute at re-prefill caps of 0, 2, 6 (the default), 30 and none (no cap): 0.58, 0.61, 0.64, 0.63 and 0.63.}
  \label{fig:eval-knobs}
\end{figure}

\label{sec:eval-abl-efficiency}

\subsection{Parameter Sensitivity}
\label{sec:eval-params}

We evaluate the sensitivity of \sys to each parameter using the workload in \S\ref{sec:eval-ablation}. We vary one parameter at a time while keeping the others at their default settings.

\subsubsection{Selection: Grace Window}
\label{sec:eval-knobs-selection}

The grace window delays the removal of superseded context for a configurable number of rounds. This gives the agent time to complete actions that may still depend on earlier outputs, such as verifying an edit (\S\ref{sec:tracking}). We compare windows by task accuracy at High, where \sys commits the most context. The default 2-round window achieves the highest accuracy at 88\%, 2 percentage points above no eviction (Fig.~\ref{fig:eval-knobs}(a)). Shorter windows reduce accuracy by 2--4 points, as they may remove context before the agent finishes using it. Longer windows reduce accuracy by 6--7 points because they retain stale context for more rounds.

\subsubsection{Pressure Gate: Threshold}
\label{sec:eval-knobs-pressure}

The pressure threshold determines when \sys begins committing removals that invalidate cached prefixes. We compare thresholds, and re-prefill caps below, by passing sessions per minute, calculated as throughput under SLO multiplied by task accuracy at High. As shown in Fig.~\ref{fig:eval-knobs}(b), the default threshold of 80\% achieves 5--6\% more passing sessions per minute than thresholds of 70\% and 90\%, and 27\% more than no eviction. Throughput under SLO differs by only about 1\% across the three thresholds because they remove similar fractions of the prompt near peak load. The 80\% threshold achieves 4 percentage points higher accuracy, accounting for its higher passing-session rate.

\subsubsection{Cache-Aware Commits: Re-Prefill Cap}
\label{sec:eval-knobs-cap}
The re-prefill cap limits the number of cached tokens invalidated by each pressure-driven commit relative to the number of context tokens removed. The default cap of 6 achieves 10\% more passing sessions per minute than a cap of 0 and 2\% more than larger caps, as shown in Fig.~\ref{fig:eval-knobs}(c). At a cap of 0, the pressure gate cannot commit removals that invalidate cached KV, leaving only cold commits (\S\ref{sec:eval-abl-throughput}). This setting reduces accuracy by 5 percentage points relative to the default. Larger caps provide no additional passing sessions per minute and reduce accuracy by 2 points.

\subsection{Generalization to HBM Expansion and KV Offloading}
\label{sec:eval-offload}
\label{sec:eval-gpu}
\label{sec:eval-capacity}

\begin{figure}[t]
  \centering
  \includegraphics[width=\columnwidth]{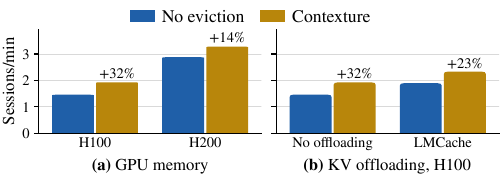}
  \vspace{-2em}
  \caption{Throughput under SLO of no eviction and \sys with Qwen MoE and Hermes on TB2.1
  : (a) on an H100 and an H200; (b) on the H100 without and with LMCache offloading to a 100\,GB host tier. \sys's gain holds with a larger HBM and with offloading.}
  \vspace{-1.em}
  \Description{Two bar charts of throughput under SLO in sessions completed per minute with Qwen MoE and Hermes on TB2.1, on one y axis from 0. (a) On an H100 with a 1.75M-token KV pool, no eviction 1.46 and Contexture 1.93 (+32\%); on an H200 with 4.44M tokens, 2.88 and 3.28 (+14\%). (b) On the H100 without offloading, 1.46 and 1.93 (+32\%); with LMCache, 1.89 and 2.33 (+23\%).}
  \label{fig:eval-capacity}
\end{figure}

A larger HBM capacity delays KV-pool saturation, while KV offloading lowers the cost of reclaiming GPU memory. We evaluate whether \sys retains its gains under both conditions. We serve Qwen MoE with Hermes on an H100 and an H200 using the same recordings, arrival schedules, and task budgets. The two GPUs provide comparable compute but differ in KV-pool capacity, with 1.75M (4.44M) tokens on the H100 (H200). \sys improves throughput under SLO by 32\% on the H100 and 14\% on the H200 (Fig.~\ref{fig:eval-capacity}(a)). The smaller gain on the H200 is consistent with the reduced impact of KV capacity on serving throughput. Shorter contexts reduce the KV-cache reads required during decoding, easing memory-bandwidth pressure even before the KV pool is full (\S\ref{sec:eval-abl-tail}).

KV offloading keeps evicted KV in host memory and reloads it when its session resumes, instead of prefilling it again (\S\ref{sec:limitations}). We add LMCache~v0.5.7~\cite{liu2025lmcache} to no eviction and \sys on the H100, configured per its recipe~\cite{lmcache2026qwenrecipe} with a 100\,GB host tier. Offloading raises no eviction's throughput under SLO by 29\%(Fig.~\ref{fig:eval-capacity}(b)); \sys alone reaches 2\% more. With offloading, \sys still gains 23\% (22\% with Qwen on the H200, where offloading adds 16\% to no eviction). Offloading cuts the cost of evicting KV, but a running request still holds and attends over its full context on the GPU; \sys shortens that context, so the two are composable.

\subsection{Overhead}
\label{sec:eval-overhead}

We measure \sys's request-path overhead across both models, benchmarks, and harnesses at their highest probed loads. 
Across requests, \sys adds a median delay of 30\,ms before engine execution, accounting for 0.2\% of request time at the median. Commit decisions and prompt filtering take a median of 9\,ms, while 69\% of requests also wait a median of 25\,ms for a residency probe.
The engine answers these probes between forward passes, so the median wait is shorter than one forward pass over the batched prefill and decode tokens.
The Session Tracker maintains its per-session state in host memory. 
Its host-memory footprint per tracked token is well under 1\% of the GPU KV memory required per token. 
\sys introduces no additional GPU computation beyond the re-prefill caused by its commits.

\section{Related Work}
\label{sec:related}

\noindent\textbf{Agent context management.} Harnesses shorten session histories through threshold-based summarization~\cite{anthropic2026claudecodehow,openai2026codexautocompact,openhands2026condenser,nous2026hermescompression,openclaw2026compaction}, observation masking~\cite{lindenbauer2025complexity,mason2026missing}, file-aware removal~\cite{cline2026contextmanager,sweagent2026historyprocessors,liu2026when}, or model-guided editing~\cite{xiao2026agentdiet,xu2026tokenpilot,shao2026clm,kang2026acon}. These methods do not coordinate removals with shared KV pressure and prefix residency. Model-guided editing also incurs additional model calls and can discard information the agent still needs~\cite{xiao2026agentdiet}. TokenPilot~\cite{xu2026tokenpilot} maintains a provider's prompt cache across a continuous task stream, but relies on cache hit and miss counts rather than runtime cache state. 
Most harness-integrated methods require adapting to each harness's history format and hooks~\cite{qwen2026qwencode,goose2026agent,deepseek2026harness}. Pichay~\cite{mason2026missing} performs masking through an API proxy but lacks visibility into the serving runtime's KV pool. \sys operates in the runtime, directly accesses shared KV pressure and prefix residency, and requires only a mapping from tool names to operation types for each new harness (\S\ref{sec:tracking}). It extends supersession to commands, builds, and tests, and coordinates removal timing with shared KV state.

\noindent\textbf{Model-level KV reduction.} Model-level methods keep the KV of high-attention tokens within a budget, some guided by cross-turn intent~\cite{zhang2023h2o,li2024snapkv,xiao2024efficient,cai2025rkv,liu2026practical,li2026intentkv}, fetch only the KV predicted to matter~\cite{lee2024infinigen}, or train a model to find stale history~\cite{kariyappa2026sidequest,zhang2026memory,sun2026scaling,wang2026swe}. They approximate attention or need model-specific training. \sys is model-agnostic: it needs no training and keeps exact attention over the context it retains. KV quantization~\cite{liu2024kivi,hooper2024kvquant} is orthogonal: it shrinks each token's KV rather than the number of tokens.

\noindent\textbf{Runtime KV management.} Serving runtimes reuse cached prefixes~\cite{kwon2023efficient,zheng2024sglang}, offload KV to host memory, SSDs or remote stores~\cite{gao2024cachedattention,yu2025pensieve,qin2025mooncake,liu2025lmcache,feng2025evicpress}, evict cached blocks by predicted reuse or recompute cost~\cite{fang2026saecache,shi2026asymcache,zheng2026pbkv,pan2025marconi}, and retain or prefetch a session's KV across tool calls and workflow steps~\cite{abhyankar2024infercept,li2025continuum,pan2025kvflow,zhang2026policy}. They decide where KV resides but leave the history unchanged, so states needed again must be reloaded or recomputed. \sys removes superseded history, which lowers the KV that later requests need and that memory tiers must store or move, and it uses the pool's pressure and cached prefixes to decide when to commit each removal. Autoscalers (ServerlessLLM~\cite{fu2024serverlessllm}, BlitzScale~\cite{zhang2026blitzscale}) scale the aggregate serving capacity, while load balancers (Llumnix~\cite{sun2024llumnix}) distribute sessions across instances to manage KV load and cache locality. \sys complements these mechanisms by reducing the KV demand generated by each session.

\noindent\textbf{Efficient context edits.} Other work makes an edit's re-prefill cheaper. Prompt caching and span removal from cached KV~\cite{gim2024prompt,yao2025cacheblend,hu2025epic,ma2026leyline,fu2026cache,shao2026clm} reuse states computed under a different preceding context. Leyline~\cite{ma2026leyline}, for example, deliberately keeps downstream states influenced by the removed content, so its result differs by design from re-prefilling the edited prompt. These implementations also constrain the model or the engine: Leyline's splice kernel~\cite{ma2026leyline} supports only MLA attention, and AKV~\cite{fu2026cache}, concurrent with this work, uses one-token KV pages in its evaluated implementation and does not yet support the linear-attention layers of the models we serve. SmoothAgent~\cite{pan2026smoothagent} keeps edits exact by scheduling their re-prefill ahead of time within available latency slack, which can limit this benefit at high load. \sys selects superseded history from tool executions and uses shared KV pressure and prefix residency to decide when to commit each removal, paying an exact re-prefill. Exact execution of the retained prompt is separate from task accuracy, which depends on what is removed.

\section{Conclusion}

We present \sys, a runtime context-eviction system that identifies superseded agent history through tool operations and coordinates removal with shared KV pressure and cache residency. Across eight configurations, \sys improves throughput under SLO by 20--73\% over the strongest of no eviction, compaction, AgentDiet, and Continuum, reduces GPU energy per completed session by up to 11\%, and maintains task accuracy within 3.7 percentage points of unloaded no eviction. These benefits persist with larger HBM capacity and KV offloading, while context reduction also lowers KV-read traffic during decoding.

\begin{acks}
The authors used Claude (Anthropic) to write experiment code, plot figures, and
refine text. The authors directed this use and reviewed, verified, and take
responsibility for all code, results, and text.
\end{acks}

\newpage
\bibliographystyle{ACM-Reference-Format}
\bibliography{contexto}
\end{sloppy}
\end{document}